\documentclass[prb,showpacs,twocolumn,aps,superscriptaddress,a4paper]{revtex4-1}
\usepackage{dcolumn,amssymb,amsmath,amsfonts,graphicx,latexsym,color,braket}

\definecolor{myblue}{rgb}{0,0,0.75}

\begin{document}

\title{Discrete Lorentz symmetry and discrete time translational symmetry}

\author{Pei Wang}
\email{wangpei@zjnu.cn}
\affiliation{Department of Physics, Zhejiang Normal University, Jinhua 321004, China}

\date{\today}

\begin{abstract}
The Lorentz symmetry and the space and time translational symmetry are
fundamental symmetries of nature. Crystals are the manifestation of the
continuous space translational symmetry being spontaneously broken into a discrete one.
We argue that, following the space translational symmetry,
the continuous Lorentz symmetry should also be broken into a discrete one,
which further implies that the continuous time translational symmetry is broken into a discrete one.
We deduce all the possible discrete Lorentz and discrete time translational
symmetries in 1+1-dimensional spacetime, and show how to build a field
theory or a lattice field theory that has these symmetries.
\end{abstract}

\maketitle

\section{Introduction}

Symmetry plays an important role in modern physics. It imposes
a constraint on the physical laws and then reduces
the number of candidate theories describing nature.
Knowing the symmetry of a system is the prerequisite for building a theory of it.
For examples, the crystals are classified by the point group and the space group
according to their symmetry under rotation, translation and reflection~\cite{Marder:2010},
the Lorentz symmetry and its generalization the Poincar\'{e}
symmetry are the basic of the relativistic quantum field theory~\cite{Greiner:1996},
the discovery of the violation of parity symmetry~\cite{Lee:1956ei,Wu:1957kk} improves our understanding of weak interaction,
and the particle-hole, the time reversal and the chiral symmetry are used to classify different topological insulators
and topological superconductors~\cite{Hasan:2010}, to name just a few.

The symmetry can be spontaneously broken at low-energy states of a system.
This mechanism has been used to explain the ferromagnetic-paramagnetic phase transition,
the existence of crystals, or the origin of mass. While it is well known that
some fundamental continuous symmetries like the space translational symmetry
can be spontaneously broken into a discrete one, whether the Lorentz symmetry
has such property is unexplored.

It has been long believed that, the Lorentz symmetry which is essentially important
in high-energy physics does not play a role in solid-state physics, especially in the study of crystals,
where the low-energy effective theories completely break the Lorentz symmetry.
In crystals the continuous space translational symmetry is spontaneously
broken into a discrete one~\cite{AltlandAlexander:2010uz}.
A crystal does not look the same under an arbitrary spatial translation of coordinates,
but only if the translation is along some specific direction with the distance
being an integer times of the lattice constant.
The continuous Poincar\'{e} group consists of spatial and temporal translations of arbitrary distance
and Lorentz transformations of arbitrary velocity~\cite{Tung:1985}, which is the symmetry group
of a relativistic field theory but not the symmetry group of crystals.

Disregarding the Lorentz symmetry leaves too much freedom in writing down a theory of crystals.
One may ask whether the crystals can have any if not all the Lorentz symmetry
which helps people to constrain the theories of them, in other words, whether
something can be left after the Lorentz symmetry is spontaneously broken.
In this paper we study to which extent the Lorentz symmetry may exist in a crystal,
and what is the consequence of it.
We argue that the continuous Lorentz symmetry contradicts the discrete
space translational symmetry and then cannot exist.
But a discrete Lorentz symmetry may exist, under which the physical laws
stay the same for two observers who are moving
at some specific velocities $\textbf{v}$ relative to each other.
$\textbf{v}$ can only take a sequence of universal discrete values.
The corresponding Lorentz transformations make up a discrete subgroup of the continuous Lorentz group.
We suggest that the Lagrangian of an effective theory describing crystals should be invariant under these
discrete Lorentz transformations, just as they are invariant under the discrete spatial translations.
Furthermore, the continuous time translational symmetry should also be spontaneously
broken into a discrete one to be compatible with the discrete Lorentz symmetry.
The discrete Lorentz transformations and the discrete temporal and spatial translations together make up a discrete Poincar\'{e} group.
We discuss how to build a field theory or a lattice field theory that has the discrete Poincar\'{e} symmetry.

In 2012, Wilczek et {\it al.}~\cite{Wilczek:2012jq,Shapere:2012gq} proposed a
theory about the spontaneous breaking of the continuous time translational symmetry into a discrete one.
The matter with such a broken symmetry is dubbed a "time crystal".
Whether there exist "time crystals" is still under debate up to
now~\cite{Li:2012iy,Bruno:2013gt,Watanabe:2015jh,Sacha:2015hb,
Else:2016gf,Khemani:2016gd,Zhang:2017,Choi:2017}.
Our theory also predicts the breaking of the continuous time translational symmetry.
But it should be distinguished from the previous theories of "time crystals".
In our theory, the broken time translational symmetry is a result of the principle of relativity (the Lorentz symmetry).

The discrete translational symmetry that we find can be represented
by a rectangular or a centered rectangular spacetime lattice, which keeps invariant under the discrete Lorentz transformations.
This finding could possibly be interesting in different contexts.
Some approaches to quantum gravity assumes that the spacetime is discretized instead of continuous,
and how to maintain the Lorentz symmetry on a spacetime lattice has then become an
important problem~\cite{Sorkin87,Sorkin03,Yamamoto,Livine,Rieffel00,Rieffel01}.
In the causal set theory, the spacetime is discretized into a random lattice
by the Poisson process, which keeps invariant under continuous Lorentz transformations statistically, in the sense
that one realization of the lattice has no Lorentz symmetry but the ensemble of them has~\cite{Sorkin87,Sorkin03}.
In the loop quantum gravity, the distance and the time interval are treated as
quantum operators. The spacetime is discretized in the sense that the corresponding operators
have discrete eigenvalues. And the Lorentz transformation becomes a unitary transformation
acting on the operators, with the Lorentz symmetry being explained as the
invariance of the eigenvalues under unitary transformations~\cite{Livine}.
In this paper, the Lorentz symmetry on the spacetime lattice has a different meaning.
On our spacetime lattice, the Lorentz symmetry is explicit but not statistical.
The cost is that only a discrete symmetry is left.
And the spacetime is classical and continuous with zero curvature (a Minkowski spacetime).
We do not try to quantize the spacetime, nor consider any theory of quantum gravity.
What we want to discuss is the symmetry group of the Minkowski spacetime
after a spontaneous symmetry breaking.

The paper is organized as follows. Sec.~\ref{sec:hypotheses} lists the basic hypotheses of our theory.
Sec.~\ref{sec:A} demonstrates why the continuous Lorentz symmetry and the discrete space translational symmetry
cannot coexist, and how the latter puts restrictions on the velocity in the Lorentz transformation.
In Sec.~\ref{sec:B}, we deduce the discrete Lorentz group that is compatible
with the discrete space translational symmetry.
In Sec.~\ref{sec:C}, we construct the discrete Poincar\'{e} group which includes both the translations and
the Lorentz transformations. We then present a corollary of our theory -
the continuous time translational symmetry is broken into a discrete one.
Sec.~\ref{sec:invariance} continues to discuss the discrete spacetime translational symmetry,
and shows how its representation (the spacetime lattice) keeps invariance under discrete Lorentz transformations.
We also discuss how to understand the time dilation and space contraction on the lattice.
Sec.~\ref{sec:causality} shows the causality between events on the spacetime lattice.
In Sec.~\ref{sec:D}, we construct the field theory that has the discrete Poincar\'{e} symmetry.
The possible features and problems when quantizing this theory are discussed in Sec.~\ref{sec:conservation}.
Sec.~\ref{sec:E} is the conclusion and outlook.

\section{Hypotheses of the theory}
\label{sec:hypotheses}

Let us recall the continuous Lorentz and Poincar\'{e} symmetry.
According to the principle of relativity in special relativity~\cite{Resnick:1968}, the physical laws
must stay the same for the observers in different reference frames
which are moving at a constant velocity relative to each other.
The transformation connecting the space and time coordinates of an event as
measured in different frames is the Lorentz transformation, which can be derived
from the principle of relativity and the principle of invariant light speed.
Again, the spacetime has translational symmetry, that is the physical laws
stay the same in the coordinate systems which are at rest
relative to each other but differ by a spatial or temporal translation of the origin.
The continuous Lorentz and translational transformations together
make up the Poincar\'{e} group~\cite{Tung:1985}, which is the symmetry group of a relativistic quantum field theory.

The crystals do not have the continuous space translational symmetry.
The breaking of the continuous space translational symmetry into a discrete one
signals the freezing phase transition from liquids to crystals. We imagine the whole
spacetime being occupied by a perfect crystal which is infinitely large and has no boundary,
thereafter, the spacetime has a discrete space translational symmetry for the observers living inside.
We propose three hypotheses about such a spacetime.

The first hypothesis is a weaker version of the principle of relativity.
It says that, for any observer in this spacetime, there exists another
observer moving at nonzero velocity relative to him and the physical
laws stay the same for them. The principle of relativity states that all the inertial reference frames moving at arbitrary velocity
are equivalent to each other in describing the physical laws.
Our first hypothesis is different from the principle of relativity which is found to
contradict the discrete space translational symmetry, as explained in Sec.~\ref{sec:A}.
We only suppose that there exists at least one velocity $\textbf{v}\neq 0$ so that two reference frames
are equivalent as one is moving at $\textbf{v}$ relative to the other.

In the second hypothesis, we suppose an invariant "light speed" which is denoted by
a constant $c$ in this paper. Note that $c$ is not the speed of light in vacuum
since the spacetime is occupied by a crystal. Instead, $c$ is the supremum limit
of the propagation speed of information and matter in the crystal.
Such a limit always exists, even if it may differ from the speed of light in vacuum.
We suppose that $c$ is the same in any reference frames.

The third hypothesis is that the spacetime has a discrete space translational symmetry. For example,
in the 3+1-dimensional spacetime, there exist three spatial vectors $\textbf{a}_1$, $\textbf{a}_2$ and $\textbf{a}_3$.
The physical laws keep invariant under a pure spatial translation of vector $\textbf{r}$ if and only if
$\textbf{r} = n_1 \textbf{a}_1+ n_2 \textbf{a}_2 + n_3 \textbf{a}_3$ with $n_1, n_2$ and $n_3$
being integers. Here a pure spatial translation
means that the two coordinate systems are at rest relative to each other
and their temporal coordinates keep the same.
$\textbf{a}_1$, $\textbf{a}_2$ and $\textbf{a}_3$ are in fact the three primitive vectors of a crystal~\cite{Marder:2010}.
Note that the third hypothesis not only says that the spatial translation of vector $n_1 \textbf{a}_1+ n_2 \textbf{a}_2
+ n_3 \textbf{a}_3$ is a symmetry transformation, but also says that
the other spatial translations are not.

%As will be shown, the above three hypotheses imply a discrete Lorentz symmetry and a discrete
%time translational symmetry of the spacetime.

\section{Continuous Lorentz symmetry and discrete translational symmetry do not coexist}
\label{sec:A}

\subsection{3+1 dimensions}

Any $n$+1-dimensional spacetime can be equipped with the above three hypotheses.
We will consider the 3+1-dimensional spacetime in this subsection and
then turn to 1+1-dimensional spacetime in next.

We consider two coordinate systems, namely {\it Jun} and {\it Tao} for convenience.
Or one can imagine {\it Jun} and {\it Tao} as two observers who
are located at the origins of the corresponding coordinate systems, respectively.
{\it Tao} is moving at a constant velocity $\textbf{v}$ relative to {\it Jun}.
And their origins differ by a four-vector $r=(r^0, r^1, r^2, r^3)^T$. The coordinate of an event
measured by {\it Tao} is denoted by the four-vector
$y'=(y'^0, y'^1, y'^2, y'^3)^T$ and that by {\it Jun} is denoted by $y$.
Here the zeroth component of a four-vector
denotes the time coordinate and the others denote the space coordinates.
The second hypothesis says that the "light speed" is the same in any reference frames.
As is well known in special relativity~\cite{Resnick:1968}, an invariant "light speed", whatever its value is,
leads to the Lorentz transformation between
the coordinates of an event measured in different reference frames.
The relation between $y'$ and $y$ is expressed as
\begin{equation}\label{eq:generaltransformation}
y' = L_\textbf{v} y + r,
\end{equation}
where $L_\textbf{v}$ is the 4-by-4 matrix of Lorentz transformation and $r$ is the translation vector.
Note that in the expression of $L_\textbf{v}$, the speed of light in vacuum must be replaced by $c$, i.e. the speed
limit of the propagation of matter or information in crystals, since it is $c$ that is invariant in our hypothesis.
For example, if $\textbf{v}$ is along the $x$-axis with an amplitude $v$,
the corresponding Lorentz matrix should be
\begin{eqnarray}\label{eq:lorentzmatrix}
L_{\textbf{v}} = \left(
\begin{array}{cccc}
\displaystyle\frac{1}{\sqrt{1-v^2/c^2}} & \displaystyle\frac{-v/c^2}{\sqrt{1-v^2/c^2}} & 0 & 0 \\
\displaystyle\frac{-v}{\sqrt{1-v^2/c^2}} & \displaystyle\frac{1}{\sqrt{1-v^2/c^2}} & 0 & 0 \\
0 & 0 & 1 & 0 \\
0 & 0 & 0 & 1
\end{array} \right).
\end{eqnarray}

Eq.~(\ref{eq:generaltransformation}) can be reexpressed in a compact form
as $y'=\Lambda(L_\textbf{v},r) y$ where $\Lambda(L_\textbf{v},r)$
denotes the combination of a Lorentz boost and a translation.
The operator $\Lambda(L_\textbf{v},r)$ is not a matrix as $r\neq 0$. But one can still define the multiplication of $\Lambda$.
Let us suppose the third observer {\it Pei} who is moving at a velocity
$\textbf{v}'$ relative to {\it Tao} and their origins differ by $r'$.
The coordinate of the event measured by {\it Pei} is denoted by $y''$
which is $y'' = \Lambda(L_{\textbf{v}'},r') y' =  \Lambda(L_{\textbf{v}'},r')
\Lambda(L_\textbf{v},r) y$. Repeatedly applying Eq.~(\ref{eq:generaltransformation})
leads to $y''= \left( L_{\textbf{v}'} L_\textbf{v}\right) y
+ \left(L_{\textbf{v}'} r+r' \right) $, the multiplication of two $\Lambda$ operators can then be expressed as
\begin{equation}\label{eq:productrule}
\Lambda(L_{\textbf{v}'},r')
\Lambda(L_\textbf{v},r) = \Lambda(L_{\textbf{v}'} L_\textbf{v}, L_{\textbf{v}'} r+r').
\end{equation}
It is straightforward to verify that the multiplication of $\Lambda$ has the associative property.

Now let us consider the first hypothesis. According to it, the physical laws stay the same for two observers as one
is moving at some velocity $\textbf{v}$ relative to the other.
The value of $\textbf{v}$ is not given in the first hypothesis which only states that $\textbf{v}$ exists.
The corresponding Lorentz transformation $\Lambda(L_\textbf{v}, 0)$ is
an element of the symmetry group of the spacetime.
Note that the translation is absent in $\Lambda(L_\textbf{v}, 0)$
because we do not yet know which translations keep the physical laws invariant.
Let us consider two coordinate systems $K$ and $K'$, while $K'$ is moving at $\textbf{v}$ relative
to $K$. $K'$ and $K$ are then equivalent for describing the physical laws. Since they are equivalent,
$K$ is not privileged over $K'$. If $K$ is equivalent to a coordinate system moving at $\textbf{v}$
relative to itself, so must be $K'$. Therefore,
the coordinate system $K''$ that is moving at $\textbf{v}$ relative to $K'$
must also be equivalent to $K'$ and then be equivalent to $K$, while the
coordinate transformation between $K''$ and $K$ is $\Lambda^2(L_\textbf{v}, 0)$.
Furthermore, the equivalence
relation is not only transitive but also reflective.
For $K'$, its equivalent partner $K$ is moving at the velocity $-\textbf{v}$
relative to it, thereafter, one coordinate system moving at $-\textbf{v}$ relative
to another must also be equivalent to it. The above statements
can be translated into the language of group. If $\Lambda(L_\textbf{v}, 0)$
is an element of any symmetry group (not necessarily the Poincar\'{e} group),
then $\Lambda^2(L_\textbf{v}, 0)$ and $\Lambda^{-1}(L_\textbf{v}, 0)$ must also
be the elements of the symmetry group, due to the property of a group.
In fact, $\Lambda^j(L_\textbf{v}, 0)$ for arbitrary integer $j$ must be an element of the group.
By using the product rule Eq.~(\ref{eq:productrule}),
we obtain $\Lambda^2(L_\textbf{v}, 0)=\Lambda(L^2_\textbf{v}, 0)$
and $\Lambda^{-1}(L_\textbf{v}, 0) = \Lambda(L^{-1}_\textbf{v}, 0) = \Lambda(L_{-\textbf{v}}, 0)$.

The third hypothesis can also be expressed in the language of group.
A pure translation between two coordinate systems can be denoted as $\Lambda(1, r)$ where
$1$ is the identity matrix which is the Lorentz transformation
between two coordinate systems being at rest relative to each other.
We distinguish the temporal and spatial components of the four-vector $r$
by expressing it as $r=(r^0, \textbf{r})^T$ with $\textbf{r}$ denoting a
three-dimensional spatial vector. $\Lambda(1, r)$ with $r=(0, \textbf{r})^T$ represents
a pure spatial translation. The third hypothesis in fact says that $\Lambda(1, r)$ is an element of the symmetry group
if and only if $\textbf{r}=n_1 \textbf{a}_1+ n_2 \textbf{a}_2 + n_3 \textbf{a}_3$.
Since the product of two translations is $\Lambda(1, r)\Lambda(1, r')=\Lambda(1, r+r')$,
the spatial translations of vector $n_1 \textbf{a}_1+ n_2 \textbf{a}_2 + n_3 \textbf{a}_3$
by themselves make up a discrete group.
It must be a subgroup of the overall symmetry group of the spacetime.

Up to now, we know that the symmetry group of the spacetime
has an element $\Lambda(L_\textbf{v}, 0)$ with $\textbf{v}\neq 0$,
and its subgroup for spatial translations contains only translations of
vector $n_1 \textbf{a}_1+ n_2 \textbf{a}_2 + n_3 \textbf{a}_3$.
Surprisingly, one can deduce from these properties that $\textbf{v}$ cannot take continuous values!
In detail, let us consider seven observers (or seven coordinate systems),
namely $K_1, K_2, \cdots, K_6$ and $K_7$. $K_2$ is moving at the velocity $-\textbf{v}$
relative to $K_1$. $K_3$ differs from $K_2$ by a spatial translation of vector $\textbf{a}_1$.
$K_4$ is moving at the velocity $\textbf{v}$ relative to $K_3$. $K_5$ is
moving at the velocity $\textbf{v}$ relative to $K_4$. $K_6$ differs from $K_5$
by a translation of vector $\textbf{a}_1$. Finally, $K_7$ is moving
at the velocity $-\textbf{v}$ relative to $K_6$. Obviously, due to
the transitivity of equivalence relation, all these seven
coordinate systems are equivalent to each other. The coordinate
of an event measured by $K_1$ and $K_7$ is denoted by $y_1$ and $y_7$, respectively.
And we use the notation $a_1 = (0, \textbf{a}_1)$. The transformation from $y_1$ to $y_7$ is then
\begin{equation}
\begin{split}
 & \Lambda^{-1}(L_\textbf{v}, 0) \Lambda(1, a_1) 
 \Lambda^2(L_\textbf{v}, 0) \Lambda(1, a_1) \Lambda^{-1}(L_\textbf{v}, 0) \\
 & = \Lambda (1, L_\textbf{v} a_1 + L^{-1}_\textbf{v} a_1).
 \end{split}
\end{equation}
Obviously, $\Lambda (1, L_\textbf{v} a_1 + L^{-1}_\textbf{v} a_1)$ is a translation
and it must be an element of the symmetry group. $L_\textbf{v}$
acting on $a_1=(0,\textbf{a}_1)$ usually generates a four-vector with nonzero temporal
component. But this temporal component exactly cancels the temporal
component of $L^{-1}_\textbf{v} a_1$, so that $L_\textbf{v} a_1 + L^{-1}_\textbf{v} a_1$
is in fact a four-vector with only spatial components and then
$\Lambda (1, L_\textbf{v} a_1 + L^{-1}_\textbf{v} a_1)$ describes
a pure spatial translation with the time coordinate keeping invariant under this transformation.
To see the cancellation between the temporal components of $L_\textbf{v} a_1$ and
$L^{-1}_\textbf{v} a_1$, one can study the example of $L_\textbf{v}$ in Eq.~(\ref{eq:lorentzmatrix}).
It is straightforward to verify that $(L_\textbf{v}+ L_{-\textbf{v}} ) a_1$ has no temporal component for
any $\textbf{v}$ and $\textbf{a}_1$.

The element $\Lambda (1, L_\textbf{v} a_1 + L^{-1}_\textbf{v} a_1)$ in the symmetry
group is a pure spatial translation. But we already know that for any spatial
translation in the symmetry group the translation vector
must be $n_1 \textbf{a}_1+ n_2 \textbf{a}_2 + n_3 \textbf{a}_3$.
Therefore, we establish an equation
\begin{equation}\label{eq:discretelorentz}
L_\textbf{v} a_1 + L^{-1}_\textbf{v} a_1 = (0, n_1 \textbf{a}_1+ n_2 \textbf{a}_2 + n_3 \textbf{a}_3)^T.
\end{equation}
This equation puts strong restrictions on the velocity $\textbf{v}$.
Since $L_\textbf{v} a_1 + L^{-1}_\textbf{v} a_1$ changes continuously with $\textbf{v}$,
Eq.~(\ref{eq:discretelorentz}) indicates that $\textbf{v}$ can only take some
discrete values. The Lorentz transformation $\Lambda(L_{\textbf{v}'}, 0)$
is not an element of the symmetry group if $\textbf{v}'$ does not satisfy Eq.~(\ref{eq:discretelorentz}), otherwise,
the third hypothesis would be violated.
If two observers are equivalent to each other, i.e., the physical laws stay the same for them, their
relative velocity must be a solution of Eq.~(\ref{eq:discretelorentz}). In Eq.~(\ref{eq:discretelorentz}),
different integer arrays $(n_1, n_2, n_3)$ give different $\textbf{v}$.
Since there are infinite number of choice for $(n_1, n_2, n_3)$,
the number of solutions of Eq.~(\ref{eq:discretelorentz}) is also infinite.

The continuous Lorentz symmetry with a continuously changing $\textbf{v}$
contradicts the discrete space translational symmetry (the third hypothesis).
If there exists any Lorentz symmetry in the inhomogeneous spacetime of a crystal,
it must be a discrete symmetry.
%It is worth emphasizing that we do not yet
%prove the existence of such a discrete Lorentz symmetry up to now,
%which will be done in next section.

\subsection{1+1 dimensions}
\label{subsec:1Dconstraint}

We have shown the contradiction between the continuous Lorentz symmetry
and the discrete space translational symmetry in 3+1-dimensional spacetime.
In fact, this contradiction exists in arbitrary $n$+1-dimensional spacetime.
Next we focus on 1+1-dimensions in which the coordinate of an event
is a two-vector $y=(t,x)^T$ where $t$ and $x$ denote the time and space coordinates, respectively.
The reason that we choose 1+1 dimensions is due to its simplicity.
Especially, the Wigner rotation~\cite{Wigner:1939} is lack in 1+1 dimensions so that
we can easily construct the discrete Lorentz symmetry group.
The presence of Wigner rotation in 2+1 and 3+1 dimensions
makes the construction of the discrete Lorentz group more complicated.
Also in experiments, a one-dimensional (1D) crystal can be realized in quantum wires.
Therefore, it is reasonable to first explore the 1+1-dimensional spacetime.
The construction of the discrete Lorentz symmetry in 2+1 and 3+1 dimensions is left in future study.

Note that for 1D crystals a single real number $a$ (the lattice constant) determines
the discrete space translational symmetry.
The crystal looks the same after a spatial translation of distance $ma$ with $m$ being an arbitrary integer.
The generator of the spatial translation is $\Lambda(1,\bar {a})$ where $\bar{a}=(0,a)^T$.
The third hypothesis in 1+1-dimensional spacetime becomes that
any pure spatial translation in the symmetry group can be expressed as
$\Lambda(1, m \bar {a})$.

The speed limit $c$ is another important
constant in our theory, thereafter, it is natural to choose
$a=c=\hbar =1$ as the unit, which will be used throughout the left paper.

In 1+1 dimensions, the Lorentz matrix relating the coordinate of an event
observed in different reference frames becomes
\begin{eqnarray}\label{eq:1Dlorentzmatrix}
L_{v} = \left(
\begin{array}{cc}
\displaystyle\frac{1}{\sqrt{1-v^2}} & \displaystyle\frac{-v}{\sqrt{1-v^2}} \\
\displaystyle\frac{-v}{\sqrt{1-v^2}} & \displaystyle\frac{1}{\sqrt{1-v^2}} 
\end{array} \right),
\end{eqnarray}
where the relative velocity $v$ is a signed real number satisfying $\left|v\right|\leq 1$
since we already set $c=1$ to the unit of velocity.
According to the first hypothesis, there exists $v\neq 0$
so that $\Lambda (L_v, 0)$ is an element of the symmetry group,
i.e., two observers are equivalent in describing the physical laws if one is moving
at the velocity $v$ relative to the other.

$\Lambda (L_v, 0)$ and $\Lambda(1,\bar {a})$ are two elements of the symmetry group.
As same as what we did in 3+1-dimensional spacetime,
we use $\Lambda (L_v, 0)$ and $\Lambda(1,\bar {a})$ to construct a
symmetry transformation $\Lambda (1, L_{v} \bar {a} + L^{-1}_{v} \bar {a} )$,
which will help us to obtain an equation of $v$.
In detail, we suppose seven observers which are equivalent to each other. $K_2$ is moving at the velocity $-{v}$
relative to $K_1$. $K_3$ differs from $K_2$ by a spatial translation of distance ${a}$.
$K_4$ is moving at the velocity ${v}$ relative to $K_3$. $K_5$ is
moving at the velocity ${v}$ relative to $K_4$. $K_6$ differs from $K_5$
by a translation of distance ${a}$. And $K_7$ is moving
at the velocity $-{v}$ relative to $K_6$. 
The relation between the coordinates of an event observed by $K_7$ and that by $K_1$ is
$y_7 = \Lambda (1, L_{v} \bar {a} + L^{-1}_{v} \bar {a} ) y_1$.
The symmetry transformation $\Lambda (1, L_{v} \bar {a} + L^{-1}_{v} \bar {a} )$ is a pure spatial translation,
indicating that $L_{v} \bar {a} + L^{-1}_{v} \bar {a}$ must be an integer times of
$\bar{a}$. We then obtain
\begin{eqnarray}\label{eq:1Ddiscretev}  \nonumber
L_{v} \bar {a} + L^{-1}_{v} \bar {a} = & \left( \begin{array}{c} 0 \\
\displaystyle\frac{2}{\sqrt{1-v^2}} a \end{array} \right) \\ = & m \left( \begin{array}{c}
0 \\ a \end{array} \right) ,
\end{eqnarray}
which can be further simplified into
\begin{equation}\label{eq:equivalencevelocity}
\frac{2}{\sqrt{1-v^2}} = m,
\end{equation}
where $m = 2, 3, 4, \cdots$ is an integer larger than one.
In a 1+1-dimensional spacetime that obeys our three hypotheses,
the physical laws stay the same for two observers only if
the relative velocity between them is $v= \pm \displaystyle\sqrt{1-\frac{4}{m^2}}$.
$m=2$ corresponds to $v=0$, that is two observers are at rest relative to each other.
And $\left| v \right|$ increases monotonically with $m$.
Since $m$ has no supremum limit, $v$ can take infinite number of values
even if $\left| v \right|$ cannot exceed the speed limit $c=1$.
Let us list a few possible values of $v$, which are $v=0, \pm \sqrt{5}/3, \pm \sqrt{3}/2, \cdots$.
Recall that the unit of $v$ is $c$, and then the smallest nonzero value of $\left| v\right|$ is $\sqrt{5}/3 c$.
Two observers moving at a relative speed lower than $\sqrt{5}/3 c$
are always not equivalent except that they are at rest relative to each other.
In next text, we call the relative velocity $v$ at which two observers are equivalent the {\bf equivalence velocity}.
An interesting observation is that the equivalence velocity is independent of $a$.
It is the same in 1D crystals with different lattice constants.
Once if the continuous space translational symmetry is broken into a discrete one,
no matter how small $a$ is, the equivalence velocity immediately loses its continuity.

\section{Discrete Lorentz symmetry}
\label{sec:B}

As shown in above, the 1+1-dimensional spacetime with discrete space translational symmetry
can only have a discrete Lorentz symmetry if not none at all.
Eq.~(\ref{eq:equivalencevelocity}) gives the necessary condition for the
equivalence velocity $v$ in the Lorentz transformation.
But it is not the sufficient condition. In fact, it is impossible for the observers moving at
the relative velocity $v(m)=\pm \displaystyle\sqrt{1-\frac{4}{m^2}}$ for arbitrary $m$
to be equivalent to each other. In other words, the symmetry group cannot
contain all the Lorentz transformations $\Lambda(L_{v(m)},0)$ for $m\geq 2$,
because such a set of $\Lambda(L_{v(m)},0)$ are not closed under multiplication!
To see this point, let us suppose that $\Lambda(L_{v(m)},0)$ for $m=3$ and $m=4$
are both the elements of the symmetry group. This is to say that
three observers $Jun$, $Tao$ and $Pei$ are equivalent to each other if $Tao$ is moving
at the velocity $v(3)=\sqrt{5}/3$ relative to $Jun$ and $Pei$ is moving at $v(4)=\sqrt{3}/2$ relative
to $Tao$. The Lorentz transformation relating the coordinate observed by $Jun$
to that by $Pei$ is $L_{v(4)} L_{v(3)} = L_{v'}$ where $v'$ denotes the velocity
of $Pei$ relative to $Jun$. By using the velocity-addition formula in special relativity
which can also be derived from Eq.~(\ref{eq:1Dlorentzmatrix}), we find that
$v'=\left( v(3)+v(4)\right)/\left(1+ v(3)v(4)\right)$. However, $2/\sqrt{1-v'^2}$
is not an integer, so that $v'$ cannot be an equivalence velocity because
this violates the third hypothesis. Therefore, the assumption of $v(m=3)$ and $v(m=4)$ being both
the equivalence velocity must be false.

A question then arises as to which Lorentz transformations
$ \Lambda(L_{v},0)$ with the velocity $ v=\pm \displaystyle\sqrt{1-\frac{4}{m^2}}$
can be in the symmetry group which must be closed under multiplication.
The answer appears to be simple.
First, the identity transformation at $v=0$ or $m=2$ must be an
element of the group. For the other elements, we strictly prove that (see App.~\ref{sec:app1}
for the detail) only the transformations that are generated by a single integer $g > 2$
can make up a group. All the Lorentz transformations in the group can be expressed as
$\Lambda(L_{v_j(g)},0)$ with the relative velocity being
\begin{equation}
v_j(g)= \textbf{sgn}(j) \displaystyle\sqrt{1-\displaystyle\frac{4}{m_j^2(g)}},
\end{equation}
where $\textbf{sgn}(j)$ denotes the sign of the integer $j$. And
$m_j(g)$ is an integer sequence generated by $g$. For $j\geq 0$, $m_0=2$ and $m_1=g$ are the first
two integers in the sequence, and the left ones are iteratively generated according to
\begin{equation}\label{eq:iterativemequation}
m_{j+1}= g m_j - m_{j-1}.
\end{equation}
For $j<0$, $m_j$ is obtained by using the property that it is an even function, i.e. $m_j=m_{-j}$.
The Lorentz transformations $\Lambda(L_{v_j(g)},0)$ for $j=0,\pm 1, \cdots$
make up a cyclic group - the discrete Lorentz group which is denoted as $\textbf{L}$. $\textbf{L}$
is uniquely determined by the integer $g$ which is called the generator of the group.
The closure of the group under multiplication can be proved by using the relations
$\Lambda(L_{v_j(g)},0)\Lambda(L_{v_i(g)},0) = \Lambda(L_{v_{i+j}(g)},0)$
and $\Lambda^j(L_{v_1(g)}, 0) = \Lambda(L_{v_j(g)},0)$. The velocity-addition formula
reads $v_{i+j}(g) = \left(v_i(g)+v_j(g)\right)/\left(1+v_i(g)v_j(g)\right)$.
Note that the Lorentz matrix in terms of $m_j$ is expressed as
\begin{eqnarray}\label{eq:lorentzmatrixwithm}
L_{v_j} = \left(
\begin{array}{cc}
\displaystyle {m_j}/2 & \displaystyle\frac{-\textbf{sgn}(j)}{2}\sqrt{m^2_j-4} \\
\displaystyle\frac{-\textbf{sgn}(j)}{2}\sqrt{m^2_j-4} & \displaystyle{m_j}/{2} 
\end{array} \right). \nonumber \\
\end{eqnarray}

Table~\ref{tab:msequence} enumerates the first few elements in the sequence $m_j(g)$
generated by $g=3$ or $g=4$. The corresponding equivalence velocities
are also displayed. $m_j$ increases exponentially with $j$, according to Eq.~(\ref{eq:iterativemequation}).
A more elegant expression of $m_j$ can be found in App.~\ref{sec:app2}.
\begin{table}[h]
\renewcommand\arraystretch{2.5}
\centering
         \vspace{0.1 in}
        \begin{tabular}{c c c c c c c c c c c c c c c c c c}
        \hline\hline 
        & $m_0$ && $m_1 (g)$ && $m_2$ && $m_3$ && $m_4$ && $m_5$ && $m_6$ & \\ \hline 
        & 2 && 3 && 7 && 18 && 47 && 123 && 322 & \\ \hline
        & $v_0$ && $v_1$ && $v_2$ && $v_3$ && $v_4$ && $v_5$ && $v_6$ & \\ \hline 
         & 0 && $\displaystyle\frac{\sqrt{5}}{3}$ && $\displaystyle\frac{3\sqrt{5}}{7}$
         && $\displaystyle\frac{8\sqrt{5}}{18} $
         && $\displaystyle\frac{21\sqrt{5}}{47}$ && $\displaystyle\frac{55\sqrt{5}}{123}$
         && $\displaystyle\frac{144\sqrt{5}}{322}$ & \\
        \hline\hline
	\end{tabular}

        \begin{tabular}{c c c c c c c c c c c c c c c c c c c}
        & $m_0$ && $m_1 (g)$ && $m_2$ && $m_3$ && $m_4$ && $m_5$ && $m_6$ & \\ \hline 
        & 2 && 4 && 14 && 52 && 194 && 724 && 2702 & \\ \hline
        & $v_0$ && $v_1$ && $v_2$ && $v_3$ && $v_4$ && $v_5$ && $v_6$ & \\ \hline 
         & 0 && $\displaystyle\frac{\sqrt{3}}{2}$ && $\displaystyle\frac{4\sqrt{3}}{7}$
          && $\displaystyle\frac{15\sqrt{3}}{26} $
         && $\displaystyle\frac{56\sqrt{3}}{97}$ && $\displaystyle\frac{209\sqrt{3}}{362}$ 
         && $\displaystyle\frac{780\sqrt{3}}{1351}$ &\\
        \hline\hline
	\end{tabular}
		\caption{The integer sequence $m_j(g)$ generated by $g=3$ (top) and $g=4$ (bottom).
		The corresponding equivalence velocities $v_j(g)$ are displayed below $m_j(g)$.}
		\label{tab:msequence}
\end{table}

The cyclic Lorentz groups appear to be the inevitable consequence of our three hypotheses.
In a spacetime where our three hypotheses stand, when an observer
is writing down the equations of physical laws, he knows that
the only observers who are using the same equations as him must
be those who are moving at the equivalence velocity $ \pm \sqrt{1-4/m_j(g)^2}$
relative to him. The other observers moving at different velocities have
different equations for the physical laws.
And $\Lambda(L_{v_j(g)}, 0)$ is the transformation relating the coordinate of an event
observed by him to those by the other equivalent observers.
It is worth emphasizing that each spacetime (each crystal)
has a unique generator $g$. But different spacetimes (different crystals) may have different generators.

\section{Discrete Poincar\'{e} symmetry}
\label{sec:C}

We derive Eq.~(\ref{eq:equivalencevelocity}) by studying
the coordinate transformation between seven well-designed equivalent coordinate systems,
namely $K_1, \cdots, K_6$ and $K_7$.
One may wonder whether it is possible to design some equivalent coordinate systems
that finally lead to a paradox and then falsifies our three hypotheses.
The answer is no! In fact, an overall symmetry group which includes the discrete Lorentz group
and the discrete space translational group as its subgroups does exist.
Our three hypotheses are self-consistent. We will discuss this overall symmetry group
- the discrete Poincar\'{e} group in this section.

The physical laws stay the same for two observers
if and only if the coordinate transformation between them is an element
of the symmetry group of the spacetime. Let us use $\mathcal{P}$ to denote the overall symmetry
group of the spacetime where our three hypotheses stand. According to the above discussions,
the subgroup of $\mathcal{P}$ for pure boost with no translation
must be the discrete Lorentz group $\left\{\Lambda(L_{v_j(g)}, 0)\right\}$.
And the subgroup of $\mathcal{P}$ for pure spatial translation with no boost or temporal translation
must be $\left\{\Lambda(1, m\bar{a})\right\}$ with $\bar{a}=(0,1)^T$
and $m$ being an integer (recall that $a=1$ is the unit of length).
In other words, two coordinate systems who have the same origins are equivalent
to each other in describing the physical laws if and only if they are
moving at the velocity $v_j(g)$ relative to each other.
And two observers who are at rest relative to each other and use the same clock
are equivalent if and only if the spatial distance between them is an integer times of $a=1$.
%$\mathcal{P}$ is obviously distinguished from the continuous Poincar\'{e} group which is much larger
%including the continuous Lorentz group and the continuous space translational group as its subgroups.

We prove that (see App.~\ref{sec:app2} for the detail) the group $\mathcal{P}$ exists
and is determined by the generator $g$. The elements of $\mathcal{P}$ can be
generally expressed as $\Lambda(L_{v_j(g)}, Y )$, where $L_{v_j(g)}$ is a
Lorentz matrix of velocity $v_j(g)$ with $j= 0, \pm 1, \pm 2, \cdots$ being an arbitrary integer. And $Y$
is a discrete translation of spacetime. It is a combination of spatial and temporal translations, being expressed as
\begin{eqnarray}\label{eq:vectorY}
Y = N_1 \left( \begin{array}{c} 0 \\ 1 \end{array}\right) + N_2 
 \left( \begin{array}{c} \displaystyle\frac{1}{2} {\sqrt{g^2-4}} \\ \displaystyle \frac{1}{2} {g}
 \end{array} \right),
\end{eqnarray}
where $N_1, N_2 = 0, \pm 1 ,\pm 2 ,\cdots$ are arbitrary integers. $\mathcal{P}$ can be written as
\begin{equation}\label{eq:expPoincare}
\mathcal{P} = \left\{ \Lambda\left(L_{v_j(g)}, Y_{N_1 N_2}(g) \right) \bigg| 
j , N_1, N_2 = 0, \pm 1, \pm 2, \cdots \right\}.
\end{equation}
The group $\mathcal{P}$ is a discrete subgroup of the continuous Poincar\'{e} group.
%Each spacetime has a unique $g$, and then has a unique $\mathcal{P}$.

It is easy to verify that the subgroup of $\mathcal{P}$ for pure boost is the discrete Lorentz group $\textbf{L}$.
Not only the Lorentz boost but also the translation depends on $g$.
The subgroup of $\mathcal{P}$ for pure translations includes the elements $\Lambda\left(1, Y_{N_1 N_2}(g) \right)$
which are the translations of vector $Y_{N_1 N_2}(g)$.
We use $\textbf{Y}$ to denote the translational group $\left\{\Lambda\left(1, Y_{N_1 N_2}(g) \right)
\right\}$ which is closed under multiplication since $\Lambda\left(1, Y_{N_1, N_2} \right)
\Lambda\left(1, Y_{N'_1, N'_2}\right)=\Lambda\left(1, Y_{N_1+N'_1, N_2+N'_2}\right)$.
$\textbf{Y}$ includes not only the pure spatial translation, but also the
temporal translation and the combination of spatial and temporal translations.
The vectors $Y_{N_1 N_2}(g)$ form a lattice including the origin in 1+1-dimensional spacetime,
which is the characteristic lattice of $\textbf{Y}$. This characteristic lattice is generated
by two primitive vectors: $(0,1)^T$ and $\left(\displaystyle\frac{1}{2} {\sqrt{g^2-4}},  \displaystyle \frac{1}{2} {g} \right)^T$.
The first vector $(0,1)^T$ corresponds to a minimum spatial translation,
while the second one corresponds to a combination of spatial and temporal translations
since $\displaystyle\frac{1}{2} {\sqrt{g^2-4}}$ and $\displaystyle \frac{1}{2} {g} $ are both nonzero.
%Therefore, in the characteristic lattice $\left\{Y_{N_1 N_2}\right\}$, a pure spatial vector with no temporal component
%must be an integer times of $\bar{a}=(0,1)^T$.
The subgroup of $\textbf{Y}$ for pure spatial translations is $\left\{\Lambda(1, m\bar{a})\right\}$, as we expected.

\begin{figure}
\centering
\includegraphics[width = 0.9 \linewidth]{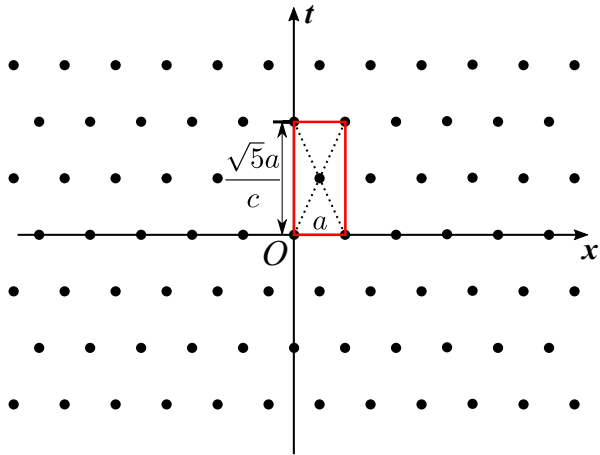}
\caption{The characteristic lattice of the translational group generated by $g=3$. The red
rectangular represents the unit cell of the lattice.}
\label{fig:g3lattice}
\end{figure}
Let us see the properties of the characteristic lattice for an odd $g$. At $g=3$, the two primitive vectors
are $(0,1)^T$ and $\left( \displaystyle\frac{\sqrt{5}}{2} ,
\displaystyle \frac{3}{2} \right)^T$. We plot the characteristic lattice of the translational symmetry for $g=3$
in Fig.~\ref{fig:g3lattice}. Note that the first component of the vector
$Y_{N_1 N_2}$ denotes the time $t$ which is the label of the vertical axis,
while the second component denotes the space $x$ which is the label of the horizontal axis.
As $g$ is an odd number, the characteristic lattice is always a centered rectangular lattice
(the red rectangular in Fig.~\ref{fig:g3lattice} represents the unit cell).
It includes vectors that lie in the direction of $t$-axis. In other words,
$\textbf{Y}$ includes pure temporal translations.

The period of the characteristic lattice in temporal and spatial directions are incommensurate. Since the unit of time
is $a/c$ according to our choice, the period of the lattice in the temporal direction is
$T=\sqrt{5} a/c$  for $g=3$. The pure temporal translation of an integer times of $\sqrt{5} a/c$
is a symmetry transformation, but those of the other
periods are not. In other words, the physical laws stay the same for two observers who are
at rest relative to each other and located at the same spot if and only if their clocks differ by an integer times of $\sqrt{5} a/c$.
Or equivalently, for a specific observer, the physical laws
change periodically with time and the period is $T=\sqrt{5} a/c$.
For a general odd generator $g$, the period of the characteristic lattice
in the temporal direction is $T=\sqrt{g^2-4} a/c$.
Following the breaking of continuous space translational symmetry,
the time translational symmetry must also be broken into a discrete one.
This is not a surprise, since the first hypothesis is a weaker version of the principle
of relativity, according to which the time cannot be separated from the space.
According to Noether's theorem~\cite{Greiner:1996}, the breaking of the continuous space (time) translational
symmetry indicates that the momentum (energy) is not conserved.
In a spacetime where our three hypotheses stands,
both the momentum and the energy are not conserved quantities.
But according to the Bloch theorem~\cite{Ashcroft:1976} or the Floquet theorem~\cite{Eckardt:2015hp}, there exist
quasi-momentum or quasi-energy which are conserved as the system has
a discrete space or time translational symmetry, respectively.

\begin{figure}
\centering
\includegraphics[width = 0.9 \linewidth]{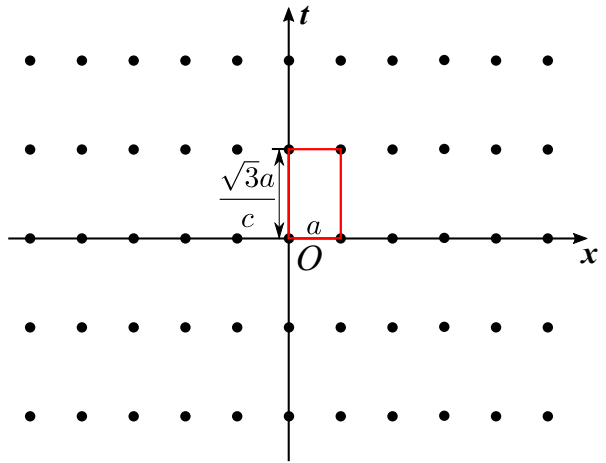}
\caption{The characteristic lattice of $\textbf{Y}$ generated by $g=4$. The red rectangular represents
the unit cell.}
\label{fig:g4lattice}
\end{figure}
The characteristic lattice for an even $g$ has a different shape.
Fig.~\ref{fig:g4lattice} plots the characteristic lattice of the translational group $\textbf{Y}$ generated by $g=4$.
%The two primitive vectors are now vertical to each other.
For an even $g$, the characteristic lattice is always a rectangular lattice.
And its period in the temporal direction is $\displaystyle {\sqrt{g^2-4}}{a}/{2c}$.

The lack of the continuous time translational symmetry and then the energy conservation law sound strange,
since it is generally believed that an isolated system should have conserved energy.
But one should not forget that our three hypotheses stand
in a spacetime where the continuous space translational symmetry
has already been spontaneously broken into a discrete one. $\mathcal{P}$ is in fact the symmetry group
of an effective theory that describes a system living
in such an inhomogeneous spacetime. Just as the electron cannot conserve its
momentum when moving within the periodic potential of crystals,
but the whole crystal as an isolated system keeps its momentum invariant.
We should understand the lack of energy conservation in the similar way as
we understand the lack of momentum conservation of an electron.
The system that the effective theory describes is not a real isolated system.

\section{Invariance of the characteristic lattice under the discrete Lorentz transformation,
time dilation and length contraction}
\label{sec:invariance}

\begin{figure}
\centering
\includegraphics[width = 0.8 \linewidth]{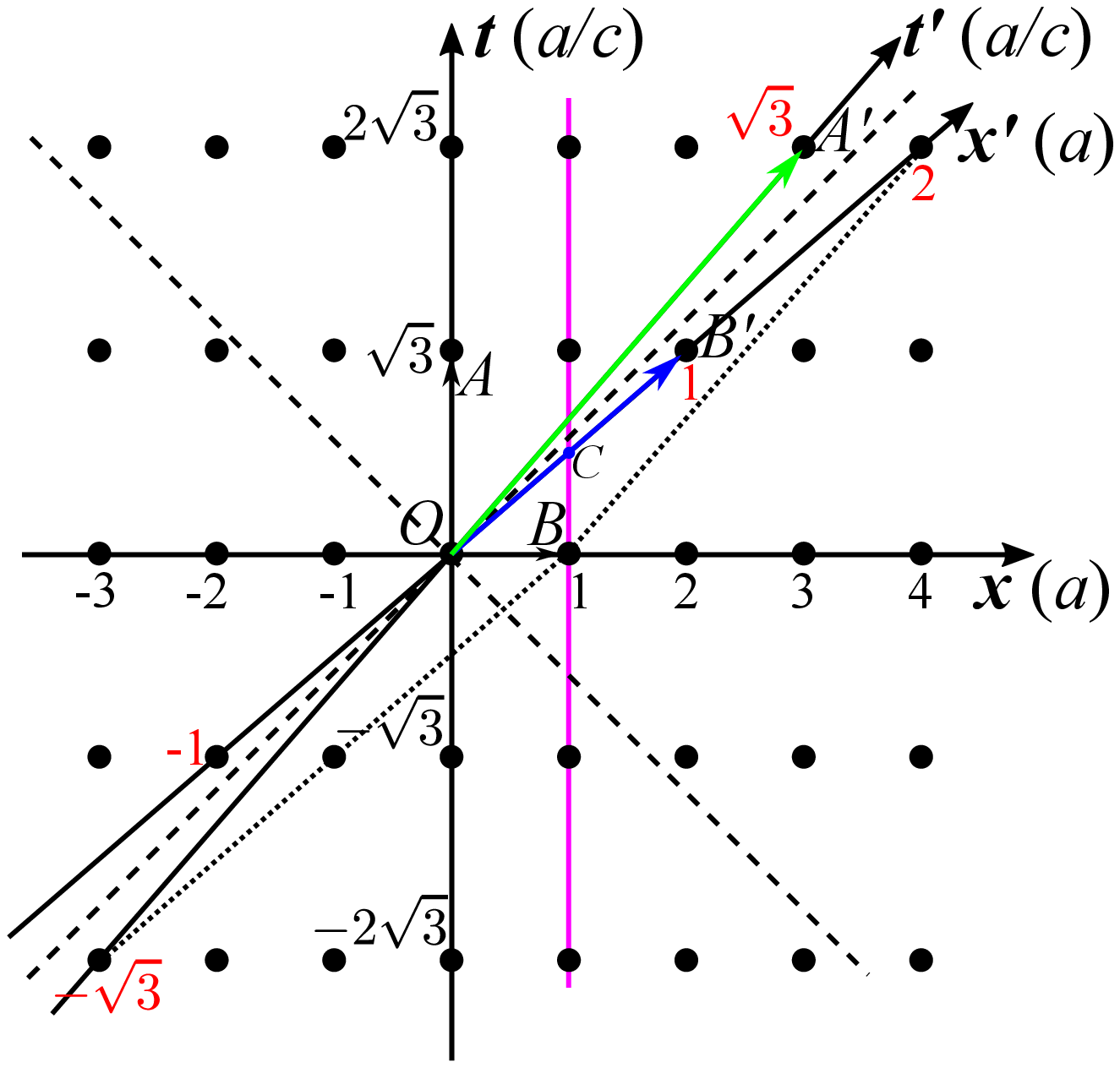}
\caption{Invariance of the characteristic lattice under the Lorentz transformation. $t'$ and $x'$ are
the time and space axes of a frame $K'$ moving at the velocity $\sqrt{3}/2 c$ relative to $K$.
The black and red numbers denote the spacetime coordinates
of an event in the frames $K$ and $K'$, respectively. The green ($\stackrel{\longrightarrow}{OA'}$)
and blue vectors ($\stackrel{\longrightarrow}{OB'}$) are the two primitive vectors of the lattice
in $K'$. The dotted lines guides the decomposition of $\stackrel{\longrightarrow}{OB}$
into $\stackrel{\longrightarrow}{OA'}$ and $\stackrel{\longrightarrow}{OB'}$.
The pink line is the world line of a particle that is static in $K$. Finally, the dashed lines represent
the light cone of $O$. }
\label{fig:invariance}
\end{figure}
We already know from above that the discrete spacetime translational symmetry can
be represented by the characteristic lattice. In this section, we show that the characteristic lattice $\{Y_{N_1 N_2} \}$ keeps
invariant under a discrete Lorentz transformation $L_{v_j}$, even if each site is transformed into another one on the same lattice.
This is expected since the spacetime translational symmetry is independent of observers.

Let us consider an observer $K$. The physical laws
for him are not the same everywhere in the continuous spacetime, but change periodically.
In next section, we will express the physical laws in terms of the Lagrangian of a field theory. We can then understand the physical
laws changing as that the coupling parameters in the theory varies in the spacetime.
Since the coupling parameters change periodically, the observer $K$ can find
a group of spacetime points which form a lattice (the characteristic lattice),
and on this lattice the coupling parameters stay the same. In other words,
the spacetime has a discrete translational symmetry represented by this lattice. Now a second observer $K'$
is moving at the velocity $v_j$ relative to $K$. What does this spacetime lattice look like in the eye of $K'$?
The answer is: exactly the same! Under the transformation from $K$ to $K'$, a site
$Y_{N_1N_2}$ on the lattice is transformed into
\begin{equation}
Y_{N'_1N'_2} = L_{v_j} Y_{N_1 N_2},
\end{equation}
which is another site on the same lattice (the proof is given in App.~\ref{sec:app2} where the relation
between the integers $N'_1$, $N'_2$, $N_1$ and $N_2$ is presented).

Fig.~\ref{fig:invariance} explains why the characteristic lattice keeps invariant under the Lorentz transformation.
We choose a spacetime with $g=4$ as an example and set $v_j=\sqrt{3}/2$ which is
the lowest positive equivalence velocity in this spacetime. The characteristic lattice of $g=4$ is
a rectangular lattice which can be seen as created by two primitive vectors $\left(0,1\right)^T$ and $\left(\sqrt{3}, 0\right)^T$,
i.e., $\stackrel{\longrightarrow}{OB}$ and $\stackrel{\longrightarrow}{OA}$ in Fig.~\ref{fig:invariance}, respectively.
Every lattice site can be expressed as $\left( n_1 \stackrel{\longrightarrow}{OB} + n_2 \stackrel{\longrightarrow}{OA}\right)$ with $n_1, n_2$
being arbitrary integers. Notice that there are infinite ways of choosing primitive vectors. For the observer $K'$,
the time and space axes are oriented in different directions, denoted by $t'$ and $x'$, respectively.
It is not an accident that $t'$ and $x'$ cross not only the origin but also some other points ($A'$ and $B'$) on the lattice.
$\stackrel{\longrightarrow}{OA'}$ (the green vector) and $\stackrel{\longrightarrow}{OB'}$ (the blue vector)
can be seen as a new pair of primitive vectors of the lattice, that is every lattice site can also be uniquely
expressed as $\left(n'_1 \stackrel{\longrightarrow}{OB'} + n'_2 \stackrel{\longrightarrow}{OA'}\right)$ with two new integers
$n'_1$ and $n'_2$. For example, we have $\stackrel{\longrightarrow}{OB}=2\stackrel{\longrightarrow}{OB'}
- \stackrel{\longrightarrow}{OA'}$ and $\stackrel{\longrightarrow}{OA}=-3 \stackrel{\longrightarrow}{OB'}+ 
2 \stackrel{\longrightarrow}{OA'}$. And in the reference frame $K'$, the length of $\stackrel{\longrightarrow}{OB'}$
and $\stackrel{\longrightarrow}{OA'}$ is $a$ and $\sqrt{3}a/c$, respectively, as same as the length
of $\stackrel{\longrightarrow}{OB}$ and $\stackrel{\longrightarrow}{OA}$ in the $K$ reference frame, respectively.
Therefore, in the eye of $K'$, the characteristic lattice is exactly the same rectangular lattice as that in the eye of $K$.

Now let us discuss the time dilation and the length contraction. Imagine a clock staying at rest relative to $K'$,
i.e., moving on the $t'$ axis relative to $K$.
The world line of this clock during one period is $\stackrel{\longrightarrow}{OA'}$ with
$t'_{A'}= \sqrt{3}a/c$ being the period of the time translational symmetry. But for
another clock staying at rest relative to $K$, the event $A'$
happens at the time $t_{A'} = 2\sqrt{3}a/c$. This reflects the fact that the clock
at rest runs twice as fast as the moving clock. In spite of the time dilation, a time interval of integer periods
keeps an interval of integer periods in any equivalent reference frames.

On the other hand, the length contraction seems to contradict the discrete translational symmetry at the first sight.
Let us choose two static points in the reference frame $K$, say the points $x=0$ and $x=a$. Because
$a$ is the proposed period of space translational symmetry,
the coupling parameters of physical laws always keep the same at these two points.
But in the reference frame $K'$, the distance between the two points contracts to $a/2$,
which seems to cause a paradox since $a$ is set to the shortest distance for a symmetric space translation.
There is in fact no paradox. The world lines of the two points are shown in Fig.~\ref{fig:invariance},
which are the $t$ axis and the pink line, respectively. One must remember that the
coupling parameters also change with time. Their changes are synchronized in the reference frame $K$,
but are not in $K'$. The coupling parameters at the spacetime points $O$ and $B$ are the same,
but they are different from the parameters at the point $C$. When the distance is measured in $K'$, the concerned
points are $O$ and $C$ which are simultaneous in $K'$. The spatial distance
between $O$ and $C$ is $a/2$ in $K'$, but the coupling parameters at these two points are different.
In previous literatures, researchers were used to take it for granted that the Lorentz contraction
forbids a spacetime lattice to have any Lorentz symmetry in the traditional meaning.
The above argument clarifies that the Lorentz contraction can coexist
with the discrete Lorentz symmetry on a proper spacetime lattice.

\section{Causality}
\label{sec:causality}

\begin{figure}
\centering
\includegraphics[width = 0.8 \linewidth]{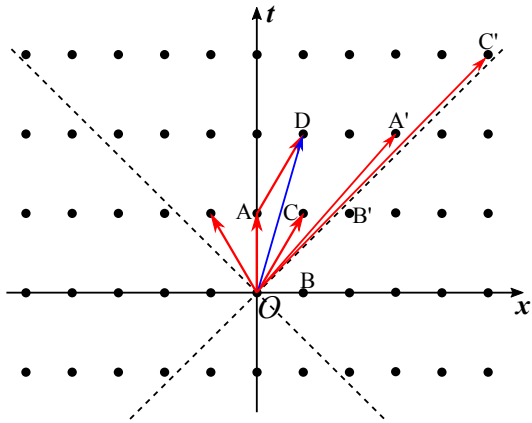}
\caption{The characteristic lattice of $g=4$ as a causal set.}
\label{fig:causality}
\end{figure}
The characteristic lattice is constituted of spacetime points (events) arranged periodically.
It is therefore interesting to discuss the causality between different events. According to the causal set theory,
the causal structure of a spacetime lattice can be used to determine the geometry of the background
manifold into which the lattice is embedded~\cite{Sorkin87}.

We use the symbol $\prec$ to denote the causal relation~\cite{Sorkin03}. $O$ and $A$ are two events. $O\prec A$
if and only if $A$ is in the future of $O$ and $A\neq O$. In Fig.~\ref{fig:causality}, the relation $O\prec A$
is represented by an arrow pointing from $O$ to $A$. In our units, the light cone of $O$
is the $45^\circ$ and $135^\circ$ lines (the dashed lines).
Therefore, an arrow is a relation if and only if it is along the positive-$t$ direction and the angle
between it and the $t$-axis is less than $45^\circ$. As is well known in special relativity,
the causal relation is invariant under the Lorentz transformation. Once if $O\prec A$ stands,
it stands in any reference frames.
There are infinite relations. One then defines the elementary relations - the links. The relation $O\prec A$ is
a link if there exists no intervening event $X$ so that $O\prec X \prec A$. The causal structure
of the spacetime lattice is totally determined by the links. The links are represented by the red arrows in Fig.~\ref{fig:causality}.
The blue arrow ($\stackrel{\longrightarrow}{OD}$) is a relation but not a link, because it can be decomposed into
$\stackrel{\longrightarrow}{OA}+\stackrel{\longrightarrow}{AD}$ which means $O\prec A \prec D$.

The characteristic lattice has periodicity. All the sites on the lattice are equivalent to each other.
Therefore, we only need to find out the links starting from a specific point, e.g., the origin $O$.
And the lattice has a mirror symmetry with respect to the $t$ axis. We then focus on the links
between the $t$-axis and the $45^\circ$ line (in the quadrant~I). One can verify that $\stackrel{\longrightarrow}{OA}$,
$\stackrel{\longrightarrow}{OC}$, $\stackrel{\longrightarrow}{OA'}$ and $\stackrel{\longrightarrow}{OC'}$ are links.
In fact, at $x=na$ for each integer $n$, there exists at most one site that is the end point of the link from $O$.
For example, $\stackrel{\longrightarrow}{OC}$ is the link pointing to $x=a$. And the arrows pointing to
all the other sites above $C$ at $x=a$ cannot be the links because they can be decomposed into $\stackrel{\longrightarrow}{OC}$
and an arrow parallel to the $t$-axis. Similarly, $\stackrel{\longrightarrow}{OA'}$
is the unique link pointing to $x=3a$. And there is no link pointing to $x=2a$ or $x=4a$.

Among $\stackrel{\longrightarrow}{OA}$, $\stackrel{\longrightarrow}{OC}$, $\stackrel{\longrightarrow}{OA'}$
and $\stackrel{\longrightarrow}{OC'}$, $\stackrel{\longrightarrow}{OA}$ and $\stackrel{\longrightarrow}{OC}$ are obviously links.
$\stackrel{\longrightarrow}{OA'}$ and $\stackrel{\longrightarrow}{OC'}$ are links because they can be
obtained by Lorentz transforming $\stackrel{\longrightarrow}{OA}$ and $\stackrel{\longrightarrow}{OC}$, respectively.
Recall that $\stackrel{\longrightarrow}{OA'}$ is the primitive vector of the lattice for the observer $K'$ (see Fig.~\ref{fig:invariance}).
Under the Lorentz transformation from $K$ to $K'$, the primitive vectors
transform as $\stackrel{\longrightarrow}{OA} \to \stackrel{\longrightarrow}{OA'}$
and $\stackrel{\longrightarrow}{OB} \to \stackrel{\longrightarrow}{OB'}$, and then
$\stackrel{\longrightarrow}{OC}= \stackrel{\longrightarrow}{OA} + \stackrel{\longrightarrow}{OB}$
transforms into $\stackrel{\longrightarrow}{OC'}=\stackrel{\longrightarrow}{OA'}+ \stackrel{\longrightarrow}{OB'}$.
The relation "link" is invariant under the discrete Lorentz transformations.
Therefore, $\stackrel{\longrightarrow}{OC'}$ and
$\stackrel{\longrightarrow}{OA'}$ must also be links.

Invariance of the relation "link" can be proved by contradiction.
Suppose that a relation $\stackrel{\longrightarrow}{OA}$ is a link in the frame $K$ but
not a link in $K'$. But $\stackrel{\longrightarrow}{OA}$ is still a causal relation in $K'$, because the causal relation
is invariant under arbitrary Lorentz transformation. We then suppose that $\stackrel{\longrightarrow}{OA}$
can be decomposed into $\stackrel{\longrightarrow}{OX} + \stackrel{\longrightarrow}{XA}$ with
$\stackrel{\longrightarrow}{OX}$ and $\stackrel{\longrightarrow}{XA}$ being the causal relations in $K'$.
But $\stackrel{\longrightarrow}{OX}$ and $\stackrel{\longrightarrow}{XA}$ must also be causal relations
in $K$. Therefore, $\stackrel{\longrightarrow}{OA}$ is also not a link in $K$, which causes a paradox.

We have infinite discrete Lorentz transformations $L_{v_j}$ with $j=1, 2, 3, \cdots$ that keep the lattice invariant.
$L_{v_j}$ for arbitrary $j$ acting on $\stackrel{\longrightarrow}{OA}$ and $\stackrel{\longrightarrow}{OC}$ produces a new pair of links.
There are then infinite possible links starting from $O$ with their end points being just above the $45^\circ$ line.

\section{Field theory that has the discrete Poincar\'{e} symmetry}
\label{sec:D}

In this section, we discuss how to construct a field theory with the discrete
Poincar\'{e} symmetry. Such a theory is expected to be the effective theory describing
a system in which the space translational symmetry is spontaneously broken into a discrete one.

We will first discuss the continuous field theory in the subsection~A. The continuous
field theory is defined in a continuous spacetime, in which the time derivative and the space derivative
of the field are present in the Lagrangian density. In the subsection~B, we turn to the lattice field theory, which
is defined on the characteristic lattice of the discrete Poincar\'{e} group. In the lattice field theory,
the coupling between different lattice sites take the place of the derivative operators. Both field theories
are invariant under the discrete Poincar\'{e} transformations.

\subsection{Field theory}
\label{sec:subfield}

Let us recall how to write down a relativistic field theory that has the continuous Poincar\'{e} symmetry.
To guarantee the Poincar\'{e} symmetry, the Lagrangian density $\mathcal{L}(y)$ must be a scalar (rank-$0$ tensor).
It is made up of constants, scalar fields, or the contraction of
higher-rank tensors. For example, the Klein-Gordon Lagrangian density for spinless particles is expressed as~\cite{Greiner:1996}
\begin{equation}\label{eq:KGLagrangian}
\mathcal{L} = \frac{1}{2} \partial^\mu \phi \partial_\mu \phi - \frac{1}{2} m^2 \phi^2,
\end{equation}
where $\phi$ is a real scalar field, $m$ is a constant, and $\partial_\mu \phi$ and $\partial^\mu \phi$ are
the covariant and contravariant vectors (rank-$1$ tensors), respectively.
Here $\mu = 0, 1$ denotes the temporal and spacial components, respectively.
The metric signature is chosen to $\left(+,-\right)$.
%The metric signature is chosen to be $\left( + , - \right)$.
%As one changes the coordinate system from $K$ to $K'$, the coordinate $y =(t,x)^T$
%transforms like $y' = \Lambda(L_v, r) y$ with $\Lambda $ being a Poincar\'{e} transformation.
%The value of $m$ in any coordinate system is the same.
%The scalar $\phi$ keeps invariant under the transformation of coordinates, satisfying $\phi(y)
%= \phi' (y')$. The contravariant vector $\partial^\mu \phi$ transforms like
%$\left(\partial^0 \phi', \partial^1 \phi'\right)^T
%= L_v \left(\partial^0 \phi, \partial^1 \phi \right)^T $, while the covariant vector $\partial_\mu \phi$
%transforms like $\left(\partial_0 \phi', \partial_1 \phi'\right)^T
%= L^{-1}_v \left(\partial_0 \phi, \partial_1 \phi \right)^T $. Note that
%$\partial_\mu \phi = \partial \phi / \partial y^\mu $ with $y^0 = t$ and $y^1 = x$ and
%$\partial_\mu \phi' = \partial \phi' / \partial y'^\mu $.
The contraction of a covariant and a contravariant vectors leaves a scalar.
Therefore, the Lagrangian density~(\ref{eq:KGLagrangian}) is a scalar which keeps invariant
under arbitrary Poincar\'{e} transformation. In other words, the Lagrangian density in two coordinate systems $K$ and $K'$
satisfies $\mathcal{L} (y) = \mathcal{L}' (y')$.

Let us see how to modify Eq.~(\ref{eq:KGLagrangian}) to obtain a new Lagrangian density
that loses the continuous Poincar\'{e} symmetry but keeps only the discrete Poincar\'{e} symmetry $\mathcal{P}$.
In other words, we want to construct a Lagrangian density that keeps invariant
under a Poincar\'{e} transformation $\Lambda$ if and only if $\Lambda$ is in $\mathcal{P}$.
There is only a single tunable parameter in the Lagrangian density~(\ref{eq:KGLagrangian}), which is the mass $m$.
Alternatively, one can treat $M = m^2$ as the tunable parameter.
To break the continuous Poincar\'{e} symmetry, we replace the constant
$M$ by a function $M(y)$. The Lagrangian density in the coordinate system $K$ becomes
\begin{equation}\label{eq:KGdiscrete}
\mathcal{L} (y) = \frac{1}{2} \partial^\mu \phi(y) \partial_\mu \phi(y) - \frac{1}{2} M(y) \phi^2(y).
\end{equation}
To obtain the Lagrangian density in a different coordinate system $K'$, we need to replace
$y$ and $\phi$ by $y'$ and $\phi'$, respectively. We obtain
$\mathcal{L}' (y') = \frac{1}{2} \partial^\mu \phi'(y') \partial_\mu \phi'(y') - \frac{1}{2} M(y') \phi'^2(y')$. Note that
$\phi$ and $\partial^\mu \phi \partial_\mu \phi$ are scalars which keep
invariant under an arbitrary Poincar\'{e} transformation, that is $\phi(y)=\phi'(y')$ and
$ \partial^\mu \phi(y) \partial_\mu \phi(y) = \partial^\mu \phi'(y') \partial_\mu \phi'(y')$.
Therefore, $\mathcal{L}' (y') = \mathcal{L} (y) $ if and only if $M(y) = M(y')$.
The Lagrangian density~(\ref{eq:KGdiscrete}) keeps invariant under a Poincar\'{e} transformation
if and only if the function $M(y)$ keeps invariant under this transformation.
In order that the Lagrangian density~(\ref{eq:KGdiscrete}) has the symmetry $\mathcal{P}$,
we must choose a function $M$ that satisfies
\begin{equation}\label{eq:Mcondition}
M(y) = M(y')
\end{equation}
where $y' = \Lambda y$ for arbitrary $\Lambda$ in $\mathcal{P}$. But $M(y) \neq M(\Lambda y)$ if $\Lambda$
is not in the group $\mathcal{P}$.

Let us see how to construct the function $M(y)$. The detailed derivation is given in App.~\ref{sec:app3}.
Here we only give the results. $M(y)$ is a periodic function in the 1+1-dimensional spacetime and has the
same periodicity as the characteristic lattice of $\mathcal{P}$.
We define two reciprocal primitive vectors which are
$k^{(1)}_\mu = \left( \displaystyle \frac{-2\pi g}{\sqrt{g^2-4}}, 2\pi \right)$ and
$k^{(2)}_\mu = \left( \displaystyle \frac{4\pi }{\sqrt{g^2-4}}, 0 \right)$.
$M$ must be expressed as the Fourier transformation
\begin{equation}\label{eq:MFourier}
M (y) = \sum_{n_1,n_2=-\infty}^\infty M_{n_1n_2} \displaystyle e^{i \left(n_1 k^{(1)}_\mu + n_2 k^{(2)}_\mu \right) y^\mu},
\end{equation}
where $n_1$ and $n_2$ are integers, $y^\mu$ is the coordinate vector
($y^0= t$ and $y^1 =x$), and $M_{n_1n_2}$ is the coefficient of the Fourier transformation.
The Einstein summation convention has been used in Eq.~(\ref{eq:MFourier}).
Furthermore, the coefficients $M_{n_1n_2}$ should satisfy
\begin{equation}\label{eq:relationM}
M_{n_1 n_2} = M_{n'_1 n'_2}
\end{equation}
for arbitrary integer pairs $(n_1, n_2)$ and $( n'_1,n'_2)$ that have the relation
\begin{eqnarray}\label{eq:n1n2pairrelation}
\bigg\{ \begin{array}{c} n_1'= z_{j+1} n_1 - z_j n_2 \\
n'_2 = z_j n_1 - z_{j-1} n_2 \end{array}.
\end{eqnarray}
Here $z_j$ is an integer sequence generated by $g$ (see App.~\ref{sec:app2} for a detailed
discussion about $z_j$). For $j\geq 0$, the first two elements
of the sequence are $z_0 = 0$ and $z_1 = 1$, and the left ones are generated according to
\begin{equation}
z_{j+1} = gz_j - z_{j-1}.
\end{equation}
And $z_{-j} = -z_j$ is an odd function of $j$. According to the properties of $z_j$,
all the integer pairs $(n_1,n_2)$ which are related to each other by Eq.~(\ref{eq:n1n2pairrelation})
make up an equivalence class. All the coefficients $M_{n_1n_2}$ with $(n_1,n_2)$ being in the same class
must be the same. But if $(n_1,n_2)$ and $(n'_1,n'_2)$ are in different classes,
$M_{n_1n_2}$ and $M_{n'_1n'_2}$ are not necessarily the same.
For example, as $g=3$, $(0,0)$ is by itself a class, and $(1,1), (2,1), (1,2), (5,2), (2,5), \cdots$
are in the same class, but $(1,3)$ is not in this class. Therefore, we have $M_{1,1}= M_{2,1}=M_{1,2}
= M_{5,2} = M_{2,5} = \cdots$, but $M_{0,0}$, $M_{1,1}$ and $M_{1,3}$ may be different
from each other. Eq.~(\ref{eq:MFourier}),~(\ref{eq:relationM})
and~(\ref{eq:n1n2pairrelation}) provide a method of constructing any function $M(y)$ that is invariant under $\mathcal{P}$.
In order that $M(y)$ is a real function, we need further require $M^*_{n_1,n_2} = M_{-n_1,-n_2}$.
A special example of $M(y)$ can be obtained by setting $M_{n_1n_2} = \bar{M} $ to a constant, that is
$M_{n_1n_2} $ in different classes are all the same. The corresponding $M(y)$ is
\begin{equation}\label{eq:specialDiracM}
M(y) = \bar{M} \displaystyle\frac{\sqrt{g^2-4}}{2} \sum_{N_1,N_2} \delta\left(x- Y_{N_1N_2}^1 \right)
 \delta\left( t - Y_{N_1 N_2}^0 \right),
\end{equation}
where $Y_{N_1N_2}^0$ and $Y_{N_1N_2}^1$ denote the temporal and the spatial components
of the vector $Y_{N_1N_2}$ (see Eq.~(\ref{eq:vectorY})), respectively. $M(y)$
is a Dirac-$\delta$ function centered at the sites of the characteristic lattice.
In the derivation of Eq.~(\ref{eq:specialDiracM}) we used the relation $\sum_n e^{i n x}= 2\pi \sum_N \delta \left(2\pi N- x\right)$.

By using the above approach, we can change any relativistic field theory (e.g., the theory of vector fields and spinor fields)
into a theory that has the discrete Poincar\'{e} symmetry $\mathcal{P}$.
We start from a theory that keeps invariant under arbitrary continuous Poincar\'{e} transformations.
We then replace the constants (e.g., the coupling or the mass) in the theory into the functions like $M(y)$.
Since $M(y)$ keeps invariant only under $\mathcal{P}$, so is the new Lagrangian density.

The charge conjugation (C), the parity (P), and the time-reversal (T) symmetries are
frequently considered in the study of field theories. The charge conjugation
concerns the transformation in the internal space, therefore, it is independent of whether
the theory has a continuous or a discrete Poincar\'{e} symmetry. The
parity and the time-reversal symmetries concern the coordinate transformation, just like the Poincar\'{e} symmetry.
If we need the Lagrangian density~(\ref{eq:KGdiscrete}) to have the PT symmetry,
we need to impose a further constraint on the function
$M(y)$, that is $M(y)=M(-y)$. According to Eq.~(\ref{eq:MFourier}),
this can be realized by demanding $M_{n_1,n_2}=M_{-n_1,-n_2}$. The
function~(\ref{eq:specialDiracM}) has this property. Because it is centered
on the characteristic lattice, and the characteristic lattice for arbitrary $g$
has the PT symmetry (see Fig.~\ref{fig:g3lattice} and~\ref{fig:g4lattice}).

\subsection{Lattice field theory}
\label{sec:sublatticefield}

Sometimes we hope to simulate the field theory by using computers and then need to discretize the spacetime.
It is impossible to discretize a spacetime without breaking the continuous Poincar\'{e} symmetry.
But one can maintain the discrete Poincar\'{e} symmetry $\mathcal{P}$ of a theory when
discretizing the spacetime into the characteristic lattice of $\mathcal{P}$.
Because the characteristic lattice keeps invariant under the transformations in $\mathcal{P}$.
Next we discuss how to build a lattice field theory~\cite{Smit:2002} that has the symmetry $\mathcal{P}$.
%on the characteristic lattice such as those in Fig.~\ref{fig:g3lattice} and~\ref{fig:g4lattice}.
%(see Fig.~\ref{fig:g3lattice} and~\ref{fig:g4lattice} for the examples of the characteristic lattice)

Let us write down a general noninteracting lattice model whose action is
\begin{equation}\label{eq:actiondiscrete}
S= \sum_{P_1 P_2 Q_1 Q_2} h_{P_1 P_2, Q_1 Q_2} \times \phi_{P_1 P_2} \times \phi_{Q_1 Q_2},
\end{equation}
where $\phi_{P_1P_2}$ denotes the value of the field $\phi$ at the lattice site $Y_{P_1 P_2}$,
and $h_{P_1 P_2, Q_1 Q_2}$ denotes the coupling between the sites $Y_{P_1P_2}$ and $Y_{Q_1 Q_2}$
with $P_1, P_2, Q_1$ and $Q_2$ being all integers. The characteristic lattice by itself
is invariant under an arbitrary transformation $\Lambda \in \mathcal{P}$. But each lattice site
changes into another one after the transformation.
Suppose that, under a transformation $\Lambda$, the sites $Y_{P_1 P_2}$
and $Y_{Q_1Q_2}$ change into $Y_{P'_1P'_2}$ and $Y_{Q'_1Q'_2}$, respectively, i.e. $Y_{P'_1 P'_2} = \Lambda Y_{P_1 P_2}$
and $Y_{Q'_1 Q'_2} = \Lambda Y_{Q_1 Q_2}$. The action $S$ keeps invariant under the transformation $\Lambda$
if and only if
\begin{equation}
h_{P_1 P_2, Q_1 Q_2} =h_{P'_1 P'_2, Q'_1 Q'_2}.
\end{equation}
For $S$ being invariant under the group $\mathcal{P}$, we
need the coupling function $h$ to be invariant under $\mathcal{P}$. As proved in App.~\ref{sec:app3},
such a coupling function depends only upon the difference
between the sites $Y_{P_1 P_2}$ and $Y_{Q_1 Q_2}$, i.e.
\begin{equation}\label{eq:discretetheorysite}
h_{P_1 P_2, Q_1 Q_2} = h(P_1-Q_1, P_2-Q_2).
\end{equation}
Furthermore, if we use the notation $Y_{N_1N_2}= Y_{P_1P_2} - Y_{Q_1Q_2}$, or equivalently, $N_1=P_1-Q_1$ and $N_2=P_2-Q_2$,
the function $h(N_1,N_2)$ must satisfy
\begin{equation}\label{eq:hlorentz}
h(N_1,N_2) = h(N'_1 , N'_2)
\end{equation}
for arbitrary integer pairs $(N_1,N_2)$ and $(N'_1 , N'_2)$ that have the relation
\begin{eqnarray}\label{eq:lorentzsite}
\bigg\{ \begin{array}{c} N'_1 = z_{j+1} N_1 + z_j N_2\\ N'_2 =
-z_j N_1 - z_{j-1} N_2  \end{array}.
\end{eqnarray}
Here the integer sequence $z_j$ appears again. Similarly, all the integer pairs $(N_1,N_2)$
which are related to each other by Eq.~(\ref{eq:lorentzsite}) make up an equivalence class.
And Eq.~(\ref{eq:hlorentz}) in fact states that $h(N_1,N_2)$ in each class has a unique value.
One should notice the difference between Eq.~(\ref{eq:lorentzsite}) and~(\ref{eq:n1n2pairrelation}).
As $(N_1,N_2)$ denotes a site on the characteristic lattice, $(n_1,n_2)$ denotes
a site on the reciprocal lattice.

Eq.~(\ref{eq:discretetheorysite}),~(\ref{eq:hlorentz}) and~(\ref{eq:lorentzsite})
give the conditions of the coupling function $h$. Equipped with a coupling function satisfying
these conditions, the lattice field theory must have the symmetry $\mathcal{P}$.

\section{Conservation of quasi-energy and quasi-momentum}
\label{sec:conservation}

We will not quantize the field theories in this paper, which is left for future study.
But it is interesting to discuss some possible features of the quantized theory. Eq.~(\ref{eq:KGdiscrete})
describes a physical system in a periodic potential which varies both with time and space.
The consequence of such a potential is well known to condensed matter community.
The electrons moving in a crystal feel a typical periodic potential varying with space.
While the electrons in an irradiated material are often treated as moving in the time-periodic
electromagnetic potential.

The momentum and energy are not conserved in the presence of periodic potentials.
They are not good quantum numbers any more.
But according to the Floquet theorem and the Bloch theorem, the quasi-energy and the quasi-momentum
are good quantum numbers instead, which are defined as the energy or the momentum modulo
$2\pi \hbar / T$ or $2\pi \hbar / a$,  respectively, where $T$ and $a$ are the periods of the potential in the temporal and spatial directions,
respectively. Our analysis in Sec.~\ref{sec:C} has already established the relation between $T$ and $a$.

Due to the spatial periodic potential, the single-particle spectrum should include a series of Bloch bands.
The energy levels are distinguished by the quasi-momentum and the band label.
If we further consider the potential being temporally periodic, the solutions of the Schr\"{o}dinger
equation must be further distinguished by the quasi-energy and the Floquet band label.
The quasi-energy, the quasi-momentum, the Bloch band and the Floquet band together determine a solution.

Recent study showed that an isolated generic Floquet system will be heated up until it
reaches the infinite-temperature state~\cite{Moessner}. But we should not forget that
our three hypotheses stand in a system where the continuous translational symmetry has
been spontaneously broken. The Lagrangian~(\ref{eq:KGdiscrete}) is the effective theory
for the symmetry-breaking state, describing something like the electrons moving in a crystal.
Therefore, the system is in fact an open system, to which the argument in Ref.~[\onlinecite{Moessner}]
does not apply.

\section{Conclusions and outlook}
\label{sec:E}

In summary, we propose a theory about the Lorentz and Poincar\'{e} symmetries
in a spacetime with discrete space translational symmetry based on three hypotheses.
In solid-state physics, the spacetime occupied by crystals is expected to have these properties.
We describe the whole symmetries of the spacetime which include the discrete Lorentz, space translational and time translational
symmetries, and show how to construct a Lagrangian or action under these symmetries.
It is worth emphasizing that, our hypotheses and results are expected to stand in the
case of the continuous space translational symmetry being spontaneously broken but not being
broken by periodic external fields.

It is worth mentioning the difference between our hypotheses and those
hiding behind the effective models of crystals (such as the Hubbard
model~\cite{Hubbard:1963}) that were frequently used up to now.
In these effective models, the kinetic energy is either expressed as
$p^2/2m$ with $p$ and $m$ denoting the momentum and the
mass, respectively, or expressed as the hopping energy between neighbor sites
on the crystal lattice. In the language of quantum field theory, the kinetic part of
the Hamiltonian looks like $ \psi (x)\left(- \hbar^2 \partial_x^2 /2m\right) \psi(x)$ in a field theory
or $ c^\dag_i c_{i+1}$ in a lattice field theory where $ \psi(x)$ and $ c_i$
denote the field operators of particles (e.g. electrons) in the crystal. The discrete space translational
symmetry is considered, but the Lorentz symmetry
is not under consideration in these models whose Lagrangians or Hamiltonians
always change as transformed from one reference frame to the other moving
at a different velocity. The complete loss of Lorentz symmetry can be viewed as a special case
of our theory as the generator of the spacetime is $g=2$ and then the discrete Lorentz group $\textbf{L}$
contains only a single element $\Lambda (1, 0)$ (the identity element).
On the other hand, the general version of our theory assumes $g>2$, and then there exist
infinite number of Lorentz transformations with nonzero velocities in the symmetry group.
Our theory involves more symmetries than the models in previous studies.
According to our theory, the Lagrangian of a model can only take some specific form,
which the previous models do not have.

Our theory is built on three hypotheses. Recall that the first hypothesis is similar to
but weaker than the principle of relativity, while the second one states an invariant speed $c$
which represents the speed limit of the propagation of information or matter in crystals.
These two hypotheses are not deduced from any known principles. Whether they are true
should be examined by experiments. We would like to point out some results
coming out from these hypotheses that could be checked by experiments.
One is the breaking of the continuous time translational symmetry. In a spacetime with an odd generator $g$,
the time translational symmetry has a period $T=\sqrt{g^2-4} a/c$, while in that with
an even $g$ the period is $T=\displaystyle {\sqrt{g^2-4}}{a}/{2c}$.
If the spacetime has a discrete time translational symmetry,
the local observables should change periodically with time, just as they vary periodically within the space
of the crystal lattice. Let us estimate the magnitude of the temporal period $T$
which depends on $a/c$. The lattice constant of a crystal is typically at the nanoscale.
$c$ is distinguished from the light speed in vacuum but is expected to be at the same magnitude as it.
We choose $a = 1 \text{nm}$ and $c=3 \times 10^8 \text{m/s}$,
and find $a/c = 3.3 \times 10^{-18} \text{s}$. The temporal period
is only a few attoseconds (too short), which maybe explains why such a periodicity
has not been observed up to now. An alternate way is to examine
the absorption spectrum of crystals. According to the Floquet theorem
(see Ref.~[\onlinecite{Eckardt:2015hp}] for a recent review), the energy of
a time-periodic system is not conserved, and should be replaced by the quasi-energy
which has a period of $\hbar 2\pi /T$. A resonance happens between the quasi-energy levels
whose difference is an integer times of $\hbar 2\pi /T$, which may cause
a peak at the frequency $1 /T$ in the absorption spectrum of crystals.
Note that $1 /T \sim 10^{18} \text{Hz}$ is in the frequency range of X-rays.
 
It is worth mentioning that the time-periodic oscillation of
observables has been predicted in the theory of "time crystals"~\cite{Wilczek:2012jq}.
But whether there exists a "time crystal" is still under debate. It was argued
that a time-periodic oscillation of observables cannot happen in
an equilibrium state described by the Gibbs ensemble~\cite{Watanabe:2015jh}. On the other hand,
our theory indicates that the Lagrangian of a model for crystals should be time-periodic,
in which case the idea of describing the equilibrium states by Gibbs ensemble
should be reexamined since it does not put space and time on an equal footing.

Finally, we would like to mention the open problems that are expected to be solved in future.
These include the construction of the discrete Lorentz symmetry in 2+1-dimensional
and 3+1-dimensional spacetimes, and the quantization of a field theory that has the
discrete Poincar\'{e} symmetry.

\section*{Acknowledgement}
This work is supported by NSFC under Grant No.~11304280.

\appendix

\section{The discrete Lorentz group}
\label{sec:app1}

We have proved in Sec.~\ref{subsec:1Dconstraint} that the velocity of a Lorentz
transformation in the symmetry group must be $v=\pm \displaystyle\sqrt{1-\frac{4}{m^2}}$.
In other words, the symmetry group of pure Lorentz transformations is a subset of
\begin{equation}\label{eq:velocityset}
\mathcal{V}= \left\{ \Lambda(L_{v},0) \bigg| v=\pm \displaystyle\sqrt{1-\frac{4}{m^2}},
\quad m = 2, 3, 4 \cdots \right\}.
\end{equation}
In this section, we prove that such a group must be a cyclic group generated by some integer $g$, as defined in Sec.~\ref{sec:B}.
Recall that a group is closed with respect to multiplication and the Lorentz matrix is expressed as
\begin{eqnarray}\label{eq:applorentzmatrix}
L_{v} = \left(
\begin{array}{cc}
\displaystyle\frac{1}{\sqrt{1-v^2}} & \displaystyle\frac{-v}{\sqrt{1-v^2}} \\
\displaystyle\frac{-v}{\sqrt{1-v^2}} & \displaystyle\frac{1}{\sqrt{1-v^2}} 
\end{array} \right).
\end{eqnarray}
The multiplication between two Lorentz transformations is
$\Lambda(L_{v},0)\Lambda(L_{v'},0) = \Lambda(L_{v}L_{v'},0)$ where $L_v L_{v'}$
is the product of two matrices.

The simplest group that is a subset of $\mathcal{V}$ is the trivial group containing only the identity transformation 
at $v=0$ or $m=2$. It is a special cyclic group. On the other hand, according to
our first hypothesis, the symmetry group should contain at least one Lorentz transformation with $v\neq 0$.
Let us suppose that except for the identity element the symmetry group
contains an element $\Lambda(L_{v(g)},0)$ with $v(g)=\displaystyle\sqrt{1-\frac{4}{g^2}}$ for some integer $g>2$.
Note that supposing $v(g) >0$ does not lose the generality since
$\Lambda(L_{v(g)},0)$ and $\Lambda^{-1}(L_{v(g)},0) = \Lambda(L_{-v(g)},0)$
must be the elements of a group simultaneously.
Once if $\Lambda(L_{v(g)},0)$ is an element,
according to the property of a group, $\Lambda^j (L_{v(g)},0)=\Lambda(L^j_{v(g)},0)$ must
be an element of the group for arbitrary integer $j=0, \pm 1, \pm 2, \cdots$.
The physical meaning of $\Lambda^j (L_{v(g)},0)$ is clear.
We choose an observer $K_0$, and call who is moving at the velocity
$v(g)$ relative to $K_0$ the observer $K_1$. Similarly, the observer $K_{j+1}$
is moving at the velocity $v(g)$ relative to $K_j$. Therefore,
the coordinate transformation from $K_0$ to $K_j$ is
$\Lambda^j (L_{v(g)},0)$. The set $\textbf{L}= \left\{ \Lambda^j (L_{v(g)},0) \bigg|
j =0, \pm 1, \pm 2, \cdots \right\} $ satisfies all the properties of a group.
It is a cyclic group. The set of observers $\left\{ K_j \bigg| j= 0, \pm 1, \pm 2, \cdots \right\}$
are all equivalent to each other in describing the physical laws.

Next we prove that $\textbf{L}$ is a subset of $\mathcal{V}$,
i.e., any element $\Lambda^j (L_{v(g)},0)$ of $\textbf{L}$ is also in $\mathcal{V}$.
Let us denote the velocity of $K_j$ relative to $K_0$ as $v_j$.
By definition, we have $v_0(g)=0$ and $v_1(g) =v(g)=\displaystyle\sqrt{1-\frac{4}{g^2}} $,
and the Lorentz matrix relating $K_0$ to $K_j$ is $L_{v_j(g)} = L^j_{v_1(g)} $.
Similarly, the velocity of $K_j$ relative to $K_i$ is denoted as $v_{j-i}$ which satisfies $ L_{v_{j-i}(g)}=L^{j-i}_{v_1(g)} $,
and the velocity of $K_i$ relative to $K_0$ is $v_i$ satisfying $L_{v_i(g)}=L^i_{v_1(g)} $.
Note that $L^{i}_{v_1(g)}$ is the $i$th power of $L_{v_1(g)}$.
We then have $L^j_{v_1(g)} = L^i_{v_1(g)} L^{j-i}_{v_1(g)} $ or
$L_{v_j(g)} = L_{v_i(g)} L_{v_{j-i}(g)}$. By using the expression of $L_v$ in Eq.~(\ref{eq:applorentzmatrix}), we obtain
\begin{equation}\label{eq:velocityrelative}
v_{j}(g) = \displaystyle \frac{v_i(g) + v_{j-i}(g)}{1+ v_i(g)v_{j-i}(g)}.
\end{equation}
Eq.~(\ref{eq:velocityrelative}) is the velocity-addition formula which is as same as that
in special relativity, because both are derived from the Lorentz transformation.
It is easy to verify that $v_i$ is an odd function of $i$, i.e., $v_{-i}=-v_i$,
and $v_{j \pm i}=\displaystyle \frac{v_j \pm v_{i}}{1\pm v_iv_{j}}$ according to Eq.~(\ref{eq:velocityrelative}).

Now we define $m_j(g) = \displaystyle\frac{2}{\sqrt{1-v_j(g) ^2}}$ for each velocity $v_j(g)$.
One can easily see $m_0 = 2$ and $m_1(g) = g$ from $v_0=0$ and $v_1(g) =\displaystyle\sqrt{1-\frac{4}{g^2}} $, respectively.
By definition, $m_j(g) = m_{-j}(g)$ is an even function of $j$.
To prove that $\Lambda^j (L_{v(g)},0)=\Lambda(L_{v_j(g)},0)$ is an element of $\mathcal{V}$,
we only need to prove that $m_j(g)$ is an integer. This is done by finding an iterative formula for $m_j(g)$.
Expressing $m_{j\pm i}$ by using $v_{j \pm i}$ and then by $v_i$ and $v_j$, we obtain
\begin{equation}\label{eq:iterativem}
m_{i+j} + m_{j-i} = m_i m_j.
\end{equation}
Choosing $i=1$, we have
\begin{equation}\label{eq:iterativemplus1}
m_{j+1} = g m_j - m_{j-1}.
\end{equation}
Since we already know $m_0$ and $m_1$, Eq.~(\ref{eq:iterativemplus1}) can be used to
calculate $m_j$ iteratively. For example, we find $m_2 = g^2-2$, $m_3 = g^3 -3 g, \cdots $.
The numbers $m_0, m_1, m_2, \cdots$ make up an infinite sequence.
And because $m_0 = 2$ and $m_1 = g$ are both integers, $m_j$ for arbitrary $j$
in the sequence must be an integer according to Eq.~(\ref{eq:iterativemplus1}).
Therefore, $\Lambda^j (L_{v(g)},0)$ for arbitrary $j$ and $g$ is an element of $\mathcal{V}$,
and $\textbf{L}$ is a subset of $\mathcal{V}$.

Up to now, we proved that the cyclic group $\textbf{L}$ generated by an integer $g$ is a subset of $\mathcal{V}$.
Next we prove that $\textbf{L}$ is the only possible group that is a subset of $\mathcal{V}$.
We will construct a proof by contradiction.
We assume that there exists a group $\mathcal{G}$ which is included in $\mathcal{V}$
but not a cyclic group. By definition, $\mathcal{G}$ must include at least two
elements $\Lambda(L_{v(g)},0) $ and $\Lambda(L_{v(g')},0) $ where
$g, g'>2$ are both integers and not in the same sequence $m_j$ generated by an integer.
Especially, $g$ and $g'$ are not in the sequence generated by each other. Without loss of generality, we suppose $2<g<g'$.
Remember that $g'$ is not in the sequence generated by $g$, i.e. $g' \neq m_j(g)$ for arbitrary $j$.
According to the property of group, $\Lambda(L_{v''},0) = \Lambda(L_{v(g')},0)
\Lambda^{-1}(L_{v(g)},0)$ is an element of $\mathcal{G}$.
The velocity-addition formula reads $v'' = (v(g')-v(g))/(1-v(g)v(g'))$.
We can deduce $v''<v(g')$ since $v(g) > 0$, and also $v'' \neq v(g)$, otherwise, we have $v(g')=2v(g) /(1+v(g)^2)$
and then $g'  = m_2(g)$ which contradicts $g' \neq m_j(g)$ for arbitrary $j$.
Since $\Lambda(L_{v''},0)$ is in $\mathcal{G}$, it must be also in $\mathcal{V}$, thereafter,
$g'' = 2/\sqrt{1-v''^2}$ is an integer satisfying $g'' \neq g$ and $g'' < g'$, deduced from $v'' \neq v(g)$ and $v'' < v(g')$.
We can also deduce $g''>2$, since $g''=2$ indicates $v''=0$ and then $v(g)=v(g')$ or $g=g'$ which contradicts our assumption.
In consequence, we constructed an integer $g''$ which is different
from both $g$ and $g'$ and is less than the max of them.
In the same way, we can use $g''$ and the smaller one of $g$ and $g'$ ($g$ in this case)
to construct a new integer $g'''$ that is different from $g$ or $g''$ and less than the max of them.
We can do this because $g$ ($g''$) is not in the sequence $m_j$ generated by $g''$ ($g$), otherwise,
we can deduce that $g'$ is also in the sequence $m_j$ generated by $g''$ ($g$) which contradicts the
assumption that $g$ and $g'$ cannot be in the same sequence.
The process of constructing new integers can be repeated for infinite number of times.
Every time we choose the smallest two in the integers that we already obtained to construct a new one.
The sequence of integers ($g, g'', g''', \cdots$) that we obtain are all different to each other and
all less than $g'$ and larger than $2$. But this is impossible, because
there only exist finite number of integers between $2$ and $g'$. Our assumption must be false.
The only possible groups included in $\mathcal{V}$ are cyclic groups. This finishes the proof.

\section{The discrete Poincar\'{e} group}
\label{sec:app2}

Our hypotheses infer that the overall symmetry group of the spacetime
should have next properties: its subgroup for pure Lorentz transformations is
$\textbf{L} = \left\{ \Lambda\left(L_{v_j(g)}, 0\right) \right\}$, and its subgroup for pure spatial translations
is $\textbf{A} = \{ \Lambda(1, m\bar{a}) \}$ with $m$ an integer and $\bar{a}=(0,1)^T$.
In this section, we prove that the minimum group that has these properties is
\begin{equation}\label{eq:appexpPoincare}
\mathcal{P} = \left\{ \Lambda\left(L_{v_j(g)}, Y_{N_1 N_2}(g) \right) \bigg| 
j , N_1, N_2 = 0, \pm 1, \pm 2, \cdots \right\},
\end{equation}
where
\begin{eqnarray}\label{eq:appvectorY}
Y_{N_1N_2} = N_1 \left( \begin{array}{c} 0 \\ 1 \end{array}\right) + N_2 
 \left( \begin{array}{c} \displaystyle\frac{1}{2} {\sqrt{g^2-4}} \\ \displaystyle \frac{1}{2} {g}
 \end{array} \right),
\end{eqnarray}
and
\begin{eqnarray}\label{eq:applorentzmatrixwithm}
L_{v_j} = \left(
\begin{array}{cc}
\displaystyle {m_j}/2 & \displaystyle\frac{-\textbf{sgn}(j)}{2}\sqrt{m^2_j-4} \\
\displaystyle\frac{-\textbf{sgn}(j)}{2}\sqrt{m^2_j-4} & \displaystyle{m_j}/{2} 
\end{array} \right). \nonumber \\
\end{eqnarray}
And any group that has these properties must contain $\mathcal{P}$ as the subgroup.
The proof is divided into two steps. First, we prove that $\mathcal{P}$ is a group,
i.e., $\mathcal{P}$ is closed under multiplication, and $\mathcal{P}$ has the above-mentioned properties.
Second, we prove that a group that has these properties must contain $\mathcal{P}$ by proving that
each element in $\mathcal{P}$ can be expressed as a product of the elements in $\textbf{L}$ and $\textbf{A}$.

Let us list some important properties of the integer sequence $m_j$ which will be used in the proof.
The iterative formula~(\ref{eq:iterativemplus1}) can be reexpressed as
\begin{equation}\label{eq:mjiterativestep}
\begin{split}
& m_{j+1}-\displaystyle\frac{g-\sqrt{g^2-4}}{2} m_j \\ & = \displaystyle\frac{g+\sqrt{g^2-4}}{2} \left( 
m_{j}-\displaystyle\frac{g-\sqrt{g^2-4}}{2} m_{j-1} \right),
\end{split}
\end{equation}
from which we derive an expression of $m_j$:
\begin{equation}
m_j = \left( \displaystyle\frac{g-\sqrt{g^2-4}}{2} \right)^j + \left( \displaystyle\frac{g+\sqrt{g^2-4}}{2} \right)^j.
\end{equation}
For convenience of presentation, we define a new sequence
\begin{equation}
z_j = \textbf{sgn}(j) \displaystyle\frac{\sqrt{m_j^2-4}}{\sqrt{g^2-4}}.
\end{equation}
It is straightforward to prove that $z_j$ can be expressed as
\begin{equation}
\sqrt{g^2-4} z_j =   \left( \displaystyle\frac{g+\sqrt{g^2-4}}{2} \right)^j- \left( \displaystyle\frac{g-\sqrt{g^2-4}}{2} \right)^j .
\end{equation}
The iterative formula of $z_j$ is as same as that of $m_j$, being
\begin{equation}
z_{j+1} = gz_j  - z_{j-1}.
\end{equation}
The first two elements of $z_j$ are $z_0 = 0$ and $z_1 = 1$ which are both integers, thereafter,
$z_j$ must be also an integer sequence just like $m_j$! $z_j$ for arbitrary $j$ is an integer,
and $z_j = -z_{-j}$ is an odd function of $j$. The useful formulas involving $z_j$ and $m_j$ are
\begin{eqnarray}\label{eq:appmjzjrelation}
\bigg\{ \begin{array}{c} z_{j+1} = \displaystyle\frac{g}{2} z_j + \displaystyle\frac{m_j}{2} \\ \\
z_{j-1} = \displaystyle\frac{g}{2} z_j - \displaystyle\frac{m_j}{2} \end{array}.
\end{eqnarray}
And a generalized iterative formula for $z_j$ is
\begin{equation}
z_{i+j+1}= z_{i+1}z_{j+1} - z_i z_j,
\end{equation}
which can also be expressed in a matrix form as
\begin{eqnarray}\nonumber\label{eq:appzjmatrix}
 \left( \begin{array}{cc} z_{i+1} & z_i \\ -z_i & - z_{i-1} \end{array}\right)
 \left( \begin{array}{cc} z_{j+1} & z_j \\ -z_j & - z_{j-1} \end{array} \right) 
 = \left( \begin{array}{cc} z_{i+j+1} & z_{i+j} \\ -z_{i+j} & - z_{i+j-1} \end{array} \right). \\
\end{eqnarray}
Especially, by taking $i=-j$ we obtain
\begin{eqnarray}\label{eq:appzjinverse}
 \left( \begin{array}{cc} -z_{j-1} & -z_j \\ z_j & z_{j+1} \end{array}\right)
 \left( \begin{array}{cc} z_{j+1} & z_j \\ -z_j & - z_{j-1} \end{array} \right) 
 = 1.
\end{eqnarray}

\subsection{$\mathcal{P}$ is a group}

According to Eq.~(\ref{eq:productrule}), the product of arbitrary two elements in $\mathcal{P}$ is
\begin{equation}
\begin{split}
& \Lambda\left(L_{v_j},  Y_{P_1 P_2}\right) \Lambda\left(L_{v_i}, Y_{N_1 N_2} \right) \\
& =  \Lambda\left(L_{v_j} L_{v_i}, Y_{P_1 P_2} + L_{v_j} Y_{N_1 N_2} \right) \\
& =  \Lambda\left(L_{v_{i+j}}, Y_{P_1 P_2} + L_{v_j} Y_{N_1 N_2} \right) ,
\end{split}
\end{equation}
where $i, j, N_1,N_2, P_1$ and $P_2$ are all integers.
To prove that $\mathcal{P}$ is closed with respect to multiplication, we need to prove
that $\Lambda\left(L_{v_{i+j}}, Y_{P_1 P_2} + L_{v_j} Y_{N_1 N_2}  \right)$ is in $\mathcal{P}$.
This is equivalent to prove that $L_{v_j} Y_{N_1 N_2} = Y_{N'_1N'_2}$ is a vector in
the characteristic lattice for arbitrary $L_{v_j}$ and $Y_{N_1 N_2}$ in the lattice. By using the expression
of $L_{v_j}$ (see Eq.~(\ref{eq:applorentzmatrixwithm})) and Eq.~(\ref{eq:appmjzjrelation}), we obtain
\begin{eqnarray}\label{eq:applorentzsite}
\bigg\{ \begin{array}{c} N'_1 = z_{j+1} N_1 + z_j N_2\\ N'_2 =
-z_j N_1 - z_{j-1} N_2  \end{array}.
\end{eqnarray}
Since $z_j$ for arbitrary $j$ is an integer, $N'_1$ and $N'_2$ must be integers. Therefore, $Y_{N'_1N'_2}$ is a vector
in the characteristic lattice, and then $\mathcal{P}$ is closed with respect to multiplication.

In Sec.~\ref{sec:C}, we already showed that $\textbf{L}$ and $\textbf{A}$ are
the subgroups of $\mathcal{P}$ for pure Lorentz transformations and pure spatial translations, respectively.
We conclude that $\mathcal{P}$ is a group that satisfies the conditions of the overall symmetry group.

\subsection{The symmetry group cannot be smaller than $\mathcal{P}$}

In this subsection, we prove that each element in $\mathcal{P}$ can be expressed as a product of the elements
in $\textbf{L}$ and $\textbf{A}$.

Recall that the characteristic lattice $\{Y_{N_1N_2}\}$ has two primitive vectors: $Y_{1,0} =\left(0,1\right)^T $
and $Y_{0,1}=\left(\displaystyle\frac{1}{2} {\sqrt{g^2-4}}, \displaystyle \frac{1}{2} {g}\right)^T $.
By using the expression of $L_{v_{-1}}$ in terms of $g$ (see Eq.~(\ref{eq:applorentzmatrixwithm})),
we express the second primitive vector as $Y_{0,1} = L_{v_{-1}} Y_{1,0}$. We then obtain
\begin{equation}
\Lambda(1, Y_{0,1}) = \Lambda(L_{v_{-1}}, 0) \Lambda(1, Y_{1,0}) \Lambda(L_{v_{1}}, 0).
\end{equation}
$\Lambda(1, Y_{1,0})$ denotes the minimum spatial translation which
is an element of $\textbf{A}$, and $\Lambda(L_{v_{\pm 1}}, 0)$
are the elements of $\textbf{L}$. Therefore, $\Lambda(1, Y_{N_1,N_2}) = \Lambda(1, Y_{1,0})^{N_1} \Lambda(1, Y_{0,1})^{N_2} $
can be expressed as a product of the elements in $\textbf{L}$ and $\textbf{A}$ for arbitrary $N_1$ and $N_2$.

For the element $\Lambda(L_{v_j}, Y_{N'_1N'_2})$ in $\mathcal{P}$, we can factorize it into
\begin{equation}\label{eq:PtoAandL}
\Lambda(L_{v_j}, Y_{N'_1N'_2}) = \Lambda(L_{v_j}, 0) \Lambda(1, Y_{N_1N_2}),
\end{equation}
where $Y_{N'_1N'_2} = L_{v_j} Y_{N_1N_2}$ or $Y_{N_1N_2}= L_{v_{-j}}Y_{N'_1N'_2} $.
$\left(N'_1,N'_2\right)$ and $\left(N_1,N_2\right)$ satisfy the relation~(\ref{eq:applorentzsite}).
According to Eq.~(\ref{eq:appzjinverse}), this relation is invertible and its inverse is
\begin{eqnarray}
\bigg\{ \begin{array}{c} N_1 = -z_{j-1} N'_1 - z_j N'_2\\ N_2 =
z_j N'_1 + z_{j+1} N'_2  \end{array}.
\end{eqnarray}
For arbitrary $\left({N'_1,N'_2}\right)$, we can find integers $N_1$ and $N_2$
that satisfy Eq.~(\ref{eq:PtoAandL}). This means that each element in $\mathcal{P}$
can be expressed as the product of an element in $\textbf{L}$ and $\Lambda(1, Y_{N_1N_2})$.
But the latter has been proved to be a product of the elements in $\textbf{L}$ and $\textbf{A}$.
Therefore, each element in $\mathcal{P}$ can be expressed as a product of the elements in $\textbf{L}$ and $\textbf{A}$.

%\subsection{An example of the symmetry group larger than $\mathcal{P}$}
%For convenience, we use ${\mathcal{P}}'$ to denote the group that contains $\textbf{L}$ and $\textbf{A}$
%as its subgroups for pure Lorentz transformations and pure spatial translations, respectively.
%According to the above arguments, we know that ${\mathcal{P}}'$ must contain $\mathcal{P}$ as its subgroup.

\section{Theories that has the discrete Poincar\'{e} symmetry $\mathcal{P}$}
\label{sec:app3}

\subsection{The field theory}

In this subsection, we explain how to construct the function $M(y)$ which satisfies
\begin{equation}
M(y)= M(\Lambda y)
\end{equation}
for arbitrary $\Lambda \in \mathcal{P}$. In other words, $M(y)$ is invariant under $\mathcal{P}$.
Each element of $\mathcal{P}$ can be factorized into $\Lambda\left(L_{v_j(g)},
Y_{N_1 N_2}\right) = \Lambda\left(1, Y_{N_1 N_2}\right) \Lambda\left(L_{v_j(g)}, 0 \right)$
where $ \Lambda\left(1, Y_{N_1 N_2}\right)$ and $ \Lambda\left(L_{v_j(g)}, 0 \right)$ are also the elements
of $\mathcal{P}$. Therefore, $M(y)$ is invariant under $\mathcal{P}$ if and only if
$M(y)$ is invariant under the transformations $\Lambda\left(L_{v_j(g)}, 0 \right)$ and
$ \Lambda\left(1, Y_{N_1 N_2}\right)$, i.e., $M(y)$ is invariant under the
discrete Lorentz group $\textbf{L}$ and the discrete translational group $\textbf{Y}$.

We notice that $\Lambda\left(1, Y_{N_1 N_2}\right)y = y + Y_{N_1 N_2}$.
Because $M$ is invariant under $\textbf{Y}$, we obtain $M(y) =M( y + Y_{N_1 N_2})$ for arbitrary $N_1$ and $N_2$.
This means that $M$ is a periodic function in the 1+1-dimensional
spacetime, and has the same periodicity as the characteristic lattice $\{ {Y}_{N_1N_2}\}$. Such a periodic
function can be expressed as a Fourier transformation. The characteristic lattice has two primitive vectors: $Y_{1,0}$ and $Y_{0,1}$.
For convenience of presentation, in this subsection we rename them as
$Y^{(1)} = (0,1)^T$ and $Y^{(2)} =  \left(\displaystyle\frac{1}{2} {\sqrt{g^2-4}},  \displaystyle \frac{1}{2} {g} \right)^T$.
Each vector in the characteristic lattice can be expressed as $Y_{N_1N_2} = N_1 Y^{(1)}+N_2 Y^{(2)}$.
The reciprocal lattice has also two primitive vectors which are found to be
$k^{(1)} = \left( \displaystyle \frac{-2\pi g}{\sqrt{g^2-4}}, 2\pi \right)$ and
$k^{(2)} = \left( \displaystyle \frac{4\pi }{\sqrt{g^2-4}}, 0 \right)$.
The inner product between the primitive vectors of the characteristic lattice
and the reciprocal lattice satisfies $k^{(a)}\cdot Y^{(b)} = 2\pi \delta_{a,b}$
where $a,b = 1,2$ and $\delta_{a,b}$ is the Kronecker delta function.
For the momentum vector $k= n_1 k^{(1)} + n_2 k^{(2)}$ with $n_1$ and $n_2$ being integers,
we have $e^{i k \cdot y} = e^{i k \cdot \left(y+ Y_{N_1 N_2}\right)}$. $\{ e^{i k \cdot y} \}$ at different $(n_1,n_2)$
form a basis of the periodic functions on the characteristic lattice. Therefore, $M$ must be expressed as
\begin{equation}\label{eq:appFourier}
M (y) = \sum_{n_1,n_2} M_{n_1n_2} e^{i \left(n_1 k^{(1)} + n_2 k^{(2)} \right) \cdot y},
\end{equation}
where $M_{n_1n_2}$ is the coefficient of the Fourier transformation.

$M(y)$ should also be invariant under $\textbf{L}$, which imposes a constraint
on the coefficients $M_{n_1,n_2}$. Substituting Eq.~(\ref{eq:appFourier})
into the condition $M(y) = M\left(\Lambda\left(L_{v_j}, 0 \right) y \right)$, we obtain
\begin{equation}\label{eq:appMcondition}
\begin{split}
& \sum_{n_1,n_2} M_{n_1n_2} e^{i \left(n_1 k^{(1)} + n_2 k^{(2)} \right) \cdot y} \\ &
=\sum_{n_1,n_2} M_{n_1n_2} e^{i \left(n_1 k^{(1)} L_{v_j} + n_2 k^{(2)} L_{v_j} \right) \cdot y},
\end{split}
\end{equation}
where we used the properties of the inner product and $L_{v_j}^T=L_{v_j}$. We notice that
\begin{equation}
\begin{split}
n_1 k^{(1)} L_{v_j} + n_2 k^{(2)} L_{v_j} = n'_1 k^{(1)} + n'_2 k^{(2)},
\end{split}
\end{equation}
where
\begin{eqnarray}\label{eq:appn1n2condition}
\bigg\{ \begin{array}{c} n_1'= z_{j+1} n_1 - z_j n_2 \\
n'_2 = z_j n_1 - z_{j-1} n_2 \end{array}.
\end{eqnarray}
Therefore, Eq.~(\ref{eq:appMcondition}) stands if and only if the coefficients $M_{n_1n_2}$ satisfy
\begin{equation}
M_{n_1n_2} = M_{n'_1n'_2}
\end{equation}
for the integer pairs $(n_1,n_2)$ and $(n'_1,n'_2)$ that are related to each other by Eq.~(\ref{eq:appn1n2condition}).
The relation~(\ref{eq:appn1n2condition}) is in fact an equivalence relation which is reflexive, symmetric and transitive.
The reflexivity, symmetry and transitivity can be easily proved by using the
properties of $z_j$ given in Eq.~(\ref{eq:appzjmatrix}) and~(\ref{eq:appzjinverse}).
The integer pairs $(n_1,n_2)$ that are related to each other by Eq.~(\ref{eq:appn1n2condition}) form an equivalence class.
All the coefficients $M_{n_1n_2}$ with $(n_1,n_2)$ being in the same class must be the same.

\subsection{The lattice field theory}

In this subsection, we explain how to construct the coupling function $h$ in a lattice
field theory that is invariant under $\mathcal{P}$. $h$ must satisfy
\begin{equation}\label{eq:appdiscretehcondition}
h_{P_1P_2,Q_1Q_2} = h_{P'_1P'_2,Q'_1Q'_2}
\end{equation}
with $Y_{P'_1P'_2} = \Lambda Y_{P_1P_2}$ and $Y_{Q'_1Q'_2} = \Lambda Y_{Q_1Q_2}$ for arbitrary $\Lambda \in \mathcal{P}$.

Again, each element of $\mathcal{P}$ can be factorized into
$\Lambda\left(L_{v_j(g)}, Y_{N_1 N_2}\right) = \Lambda\left(1, Y_{N_1 N_2}\right) \Lambda\left(L_{v_j(g)}, 0 \right)$.
The coupling function $h$ is invariant under $\mathcal{P}$ if and only if
it is invariant under the transformations $\Lambda\left(1, Y_{N_1 N_2}\right) $ and $ \Lambda\left(L_{v_j(g)}, 0 \right)$.
Since $h$ is invariant under $\Lambda\left(1, Y_{N_1 N_2}\right) $, we have $h_{P_1,P_2,Q_1,Q_2} =
h_{P_1+N_1, P_2+N_2, Q_1+N_1, Q_2+N_2}$ for arbitrary integers $N_1$ and $N_2$. This means that $h_{P_1P_2,Q_1Q_2}$
depends only upon the difference between $(P_1,P_2)$ and $(Q_1,Q_2)$. We can then reexpress the coupling function as
\begin{equation}\label{eq:appdiscretetheorysite}
h_{P_1 P_2, Q_1 Q_2} = h(P_1-Q_1, P_2-Q_2).
\end{equation}

Let us use the notation $Y_{N_1N_2}= Y_{P_1P_2} - Y_{Q_1Q_2}$, or equivalently, $N_1=P_1-Q_1$ and $N_2=P_2-Q_2$.
The coupling function $h_{P_1 P_2, Q_1 Q_2} $ should be invariant under the Lorentz transformation $ \Lambda\left(L_{v_j}, 0 \right)$,
under which we have $Y_{P'_1P'_2} = \Lambda\left(L_{v_j}, 0 \right)
 Y_{P_1P_2}$ and $Y_{Q'_1Q'_2} = \Lambda\left(L_{v_j}, 0 \right) Y_{Q_1Q_2}$.
We then find $Y_{N'_1N'_2} =  Y_{P'_1P'_2} - Y_{Q'_1Q'_2} = L_{v_j} Y_{N_1N_2}$.
The integer pairs $(N_1,N_2)$ and $(N'_1,N'_2)$ have the next relation:
\begin{eqnarray}\label{eq:appdiscreteNrel}
\bigg\{ \begin{array}{c} N'_1 = z_{j+1} N_1 + z_j N_2\\ N'_2 =
-z_j N_1 - z_{j-1} N_2  \end{array}.
\end{eqnarray}
Substituting Eq.~(\ref{eq:appdiscretetheorysite}) into Eq.~(\ref{eq:appdiscretehcondition}), we obtain
\begin{equation}\label{eq:apphlorentz}
h(N_1,N_2) = h(N'_1 , N'_2).
\end{equation}
The coupling function must satisfy Eq.~(\ref{eq:apphlorentz}) for being invariant under $\mathcal{P}$.
Again, the integer pairs that are related to each other by Eq.~(\ref{eq:appdiscreteNrel}) form an equivalence class.
Eq.~(\ref{eq:apphlorentz}) says that $h(N_1,N_2)$ with $(N_1,N_2)$ being in the same class must be the same.

\bibliographystyle{apsrev}
\bibliography{literature}

\end{document}